\documentclass[galaxies,review,accept,oneauthor,pdftex]{mdpi}

\setitemize{parsep=6pt,itemsep=0pt,leftmargin=*,labelsep=5.5mm,align=parleft}
\setenumerate{parsep=6pt,itemsep=0pt,leftmargin=*,labelsep=5.5mm,align=parleft}
\setlist[description]{itemsep=0mm}

\usepackage{bm}
\usepackage{amsmath}
\usepackage{amssymb}
\usepackage{textcomp}

\newcommand{\bp}{\hat{\bm{p}}}
\newcommand{\overbar}[1]{\mkern 3.5mu\overline{\mkern-3.5mu#1\mkern-3.5mu}\mkern 3.5mu}
\newcommand{\fhit}{f_{\mathrm{hit}}}
\newcommand{\IRF}{\mathrm{IRF}}
\newcommand{\intlos}{\int_{\mathrm{los}(\bp)}}
\newcommand{\dom}{{\Delta\Omega}}
\newcommand{\domtot}{\dom_\mathrm{tot}}
\newcommand{\Aeff}{{A_\mathrm{eff}}}
\newcommand{\Aeffj}{{\bar{A}_{\mathrm{eff},j}}}
\newcommand{\avAeff}{{\bar{A}_{\mathrm{eff}}}}
\newcommand{\atot}{A_\mathrm{tot}}

\newcommand{\pdfE}{f_E}
\newcommand{\pdfP}{{f_{\bp}}}

\newcommand{\sv}{\langle \sigma \mathit{v} \rangle}
\newcommand{\mdm}{m_\chi}
\newcommand{\ndm}{n_\chi}
\newcommand{\Gdm}{\Gamma_\chi}

\newcommand{\taudm}{\tau_\chi}
\newcommand{\taudmnc}{\tau_\chi^{\mathrm{LL}_{95}}}
\newcommand{\dNdE}{\frac{dN_{\gamma}}{dE}}
\newcommand{\dJdp}{\frac{dJ}{d\Omega}}
\newcommand{\djdp}{\frac{d{\cal J}}{d\Omega}}

\newcommand{\Jann}{J_\mathrm{ann}}
\newcommand{\Jdec}{J_\mathrm{dec}}
\newcommand{\Jtot}{\overline{J}}

\newcommand{\dJdpann}{\frac{d\Jann}{d\Omega}}

\newcommand{\dJdpdec}{\frac{d\Jdec}{d\Omega}}
\newcommand{\Junits}{GeV$^2$ cm$^{-5}$}

\newcommand{\svunits}{cm$^3$ s$^{-1}$}

\newcommand{\bb}{b\bar{b}}
\newcommand{\uu}{u\bar{u}}
\newcommand{\dd}{d\bar{d}}
\newcommand{\ssbar}{s\bar{s}}
\newcommand{\ttbar}{t\bar{t}}
\newcommand{\WW}{W^+W^-}
\newcommand{\tautau}{\tau^+\tau^-}
\newcommand{\mumu}{\mu^+\mu^-}
\newcommand{\ee}{e^+e^-}

\newcommand{\Non}{N_\mathrm{On}}
\newcommand{\NdSph}{N_{\mathrm{dSph}}}
\newcommand{\Nmeas}{N_{\mathrm{meas}}}

\newcommand{\Tobs}{T_\mathrm{obs}}

\newcommand{\lp}{\lambda_P}
\newcommand{\lkl}{\mathcal{L}}
\newcommand{\lklgamma}{{\lkl}_\gamma}
\newcommand{\lklgammak}{{\lkl}_{\gamma,k}}
\newcommand{\lklmu}{{\lkl}_\mu}
\newcommand{\lklJ}{{\lkl}_J}

\newcommand{\hatnu}{\hat{\bm{\nu}}}

\newcommand{\data}{\bm{\mathcal{D}}}
\newcommand{\NEbins}{N_{E'}}
\newcommand{\NPbins}{N_{\bp'}}

\newcommand{\bij}{b_{ij}}

\newcommand{\Nonij}{{N_{\mathrm{On},ij}}}
\newcommand{\Noffij}{{N_{\mathrm{Off},ij}}}
\newcommand{\Nij}{{N_{ij}}}
\newcommand{\Jobs}{\Jtot_\mathrm{obs}}
\newcommand{\tauobs}{\tau_\mathrm{obs}}

\newcommand{\sigmatau}{\sigma_\tau}
\newcommand{\sigmatausys}{\sigma_{\tau,\mathrm{sys}}}
\newcommand{\sigmataustat}{\sigma_{\tau,\mathrm{stat}}}

\newcommand{\aaa}{\alpha}
\newcommand{\aaatrue}{\alpha_{\mathrm{true}}}
\newcommand{\hataaa}{\hat{\aaa}}
\newcommand{\aaadsu}{\aaa_{2.71}}
\newcommand{\aaanc}{\aaa^{\mathrm{UL}_{95}}}

\newcommand{\svnc}{\sv^{\mathrm{UL}_{95}}}
\newcommand{\barPhi}{\overbar{\Phi}}

\firstpage{1} 
\pubvolume{xx}
\issuenum{1}
\articlenumber{5}
\pubyear{2020}
\copyrightyear{2020}
\history{Received: 29 July 2019; Accepted: 24 February 2020; Published: 17 March 2020}
\Title{Gamma-Ray Dark Matter Searches in Milky Way Satellites---A
  Comparative Review of Data Analysis Methods and Current Results}

\Author{{Javier Rico} \orcidA{}}%Please carefully check the accuracy of name  and affiliation.

\AuthorNames{Javier Rico}

\address[1]{Institut de F\'isica d'Altes Energies (IFAE), The
  Barcelona Institute of Science and Technology (BIST),  08193 Barcelona, Spain; jrico@ifae.es }

\abstract{If dark matter is composed of weakly interacting particles
  with mass in the GeV-TeV range, their annihilation or decay may
  produce gamma rays that could be detected by gamma-ray telescopes.
  Observations of dwarf spheroidal satellite galaxies of the Milky Way
  (dSphs) benefit from the relatively accurate predictions of dSph
  dark matter content to produce robust constraints to the dark matter
  properties. The sensitivity of these observations for the search for
  dark matter signals can be optimized thanks to the use of advanced
  statistical techniques able to exploit the spectral and
  morphological peculiarities of the expected signal. In this paper, I
  review the status of the dark matter searches from observations of
  dSphs with the current generation of gamma-ray telescopes:
  Fermi-LAT, H.E.S.S, MAGIC, VERITAS and HAWC. I will describe in
  detail the general statistical analysis framework used by these
  instruments, putting in context the most recent experimental results
  and pointing out the most relevant differences among the different
  particular implementations. This~will facilitate the comparison of
  the current and future results, as well as their eventual
  integration in a multi-instrument and multi-target dark matter
  search.}


\begin{document}

\section{Introduction}
The existence of a dominant non-baryonic, neutral, cold matter
component in the Universe, called {dark matter}, has been
postulated in order to explain the kinematics of galaxies in galaxy
clusters~\cite{ref:zwicky33} and stars in spiral
galaxies~\cite{ref:babcock39}, as well as the power spectrum of
temperature anisotropies of the cosmic microwave
background~\cite{ref:planck2018}. In one of the most plausible and
thoroughly studied theoretical scenarios, dark matter is composed of
weakly interacting particles with mass in the range between tens of
GeV and hundreds of TeV, generically referred to as
{WIMPs}~\cite{ref:Hut77}. The Standard Model (SM) particles that could
result from WIMP annihilation or decay would hadronize, radiate and/or
decay, producing detectable stable particles such as photons,
neutrinos, proton--antiproton pairs or electron--positron
pairs~\cite{ref:Bergstrom2000}. Looking for unambiguous spectral
and/or morphological signatures of dark matter annihilation or decay
in the extra-terrestrial fluxes of those particles is usually referred
to as {indirect} dark matter searches.

Gamma rays are promising messengers to search for WIMPs. Since they
are electrically neutral, they are not deflected by magnetic fields
and point back to their production site, and therefore could be used
to determine the underlying dark matter spatial distribution. At
non-cosmological scales, gamma rays are also essentially unaffected by
energy losses, which would preserve the features expected for 
dark matter annihilation and/or decay spectra, which depend
on the values of the dark matter mass and the branching ratios to the
different annihilation/decay channels, which could thus be studied.
Finally, the gamma-ray signal intensity would depend on the
annihilation cross-section or the decay lifetime, which could
therefore be determined if we measured a signal from an astronomical
site for which we have a good estimate of its dark matter content
based on independent measurements and/or~simulations.

N-body simulations predict the formation of cold dark matter haloes in
a hierarchical clustering fashion~\cite{ref:Dubinski1991}. dSphs form
in dark matter galactic subhalos that contain enough baryonic matter
to have activated stellar formation (pure dark matter halos should
also exist, but they remain as of yet unidentified). They are
irregular satellite galaxies with mass $\sim 10^7 M_{\odot}$ and the
largest known ratios of dark to luminous matter. The extension of the
expected gamma-ray emission from the Milky Way dSphs is typically
between $\sim 0.1$--$0.5^\circ$~\cite{ref:Geringer2014}, which is of the
order of the angular resolution of most of the current-generation
gamma-ray telescopes.
 
Gamma-ray telescopes of the current generation have performed
extensive observational campaigns of dSphs in search for dark matter
signals. Along the years, gamma-ray telescopes have progressively
adopted state-of-the-art statistical analysis techniques for their
dark matter searches, optimized to exploit the particular spectral and
morphological features expected for the signal. All the instruments
have converged into a general statistical analysis framework, albeit
with some significant differences among the different implementations.
Some of these differences are unavoidable, since they are needed to
adapt the analysis to the different experimental scenarios, whereas
others rather consist in choices of conventions, approximations, or
simplifications. These latter ones include the methods for
computing the spectral and morphological models for the expected
gamma-ray signal and associated background, their use in the statistical analysis, and the
treatment of the related statistical and systematic uncertainties.
Understanding both the similarities and the differences among the
various analysis implementations is fundamental in view of meaningful
comparison and combination of the obtained results.

In this paper, I review the present status of indirect dark matter
searches with observations of dSphs with gamma-ray telescopes. In
Section~\ref{sec:fluxes} I~summarize the formalism for the computation
of the gamma-ray fluxes expected to be produced by dark matter
processes in dSphs. In Section~\ref{sec:instruments}, I~briefly
introduce the current generation of gamma-ray telescopes, their
working principles and main features. Section~\ref{sec:analysis} is
devoted to the detailed description of the common statistical data
analysis framework used by all these instruments in their search for
dark matter in dSphs. Finally, in Section~\ref{sec:results}, I perform
a critical comparison of the particular analysis implementations,
review and contextualize the latest experimental results published by
the different instruments, and show the prospects for their
near-future combination.

\section{Gamma-Ray Signals From Dark Matter Processes in dSphs}
\label{sec:fluxes}

dSphs are among the cleanest astronomical targets for indirect dark
matter searches. They are thought to be highly dominated by dark
matter (mass-to-light ratios of the order of
$10^3$~\cite{ref:Strigari2008}), and they harbor no known
astrophysical gamma-ray sources that could produce a relevant
background.  Furthermore, dSphs contain in general no significant
amount of dark gas, which allows their dark matter distribution to be
inferred with relatively good precision from the stellar motions,
enabling in turn robust predictions of the intensity of the associated
gamma-ray signals, generally within an accuracy of one order of
magnitude~\cite{ref:Geringer2014}. Finally, given how most of the
known dSphs sit on relatively clean interstellar environments (i.e.,  
out of the Galactic plane, where the particle densities, cosmic
ray fluxes and radiation fields are small), the expected
gamma-ray signal would come from well-understood prompt
processes. Secondary processes such as inverse Compton scattering of
primary or secondary electrons, or gamma-ray cascading processes
initiated by their interaction with radiation fields (hence depending
on local details of those radiation fields), can be in general ignored
when computing the gamma-ray flux expected from dark matter at
dSphs. Therefore, since flux predictions rely on relatively few
assumptions compared to other typical observational targets like
e.g., the Galactic center or clusters of galaxies, the bounds on the
WIMP properties that can be inferred from the presence or absence of a
gamma-ray signal are also relatively robust.

If WIMPs (hereafter denoted by $\chi$) concentrate with number density
$\ndm$ in a dSph, annihilating and/or decaying with a rate $\Gdm$
and an average isotropic gamma-ray spectrum $\dNdE$, then the
differential flux of gamma rays of energy $E$ observable from Earth
coming from direction $\bp$, per unit energy and solid angle $\Omega$,
is given by the following expression:
\begin{equation}
\frac{d^2\Phi}{dE d\Omega}(E,\bp) =  \frac{1}{4\pi}
\dNdE(E) \intlos dl\, \ndm(\bp,l)\, \Gdm \quad ,
\label{eq:flux}
\end{equation}
with $l$ the distance from Earth and the corresponding integral running
over the line of sight in the direction $\bp$.

As explicitly noted in Equation~(\ref{eq:flux}), $\dNdE$ contains all
the spectral dependence of the gamma-ray flux, and therefore
determines the probability density function (PDF) for the energy of
the emitted gamma rays. On the other hand, all the morphological
dependence is contained in the line-of-sight integral, which hence
determines the PDF for the gamma-ray arrival direction. Given that we
can make relatively reliable predictions about these two PDFs, they
will constitute key ingredients in the maximum-likelihood data
analysis, as we will see below in detail.

\begin{figure}[t]
\centering
\includegraphics[width=7.75 cm]{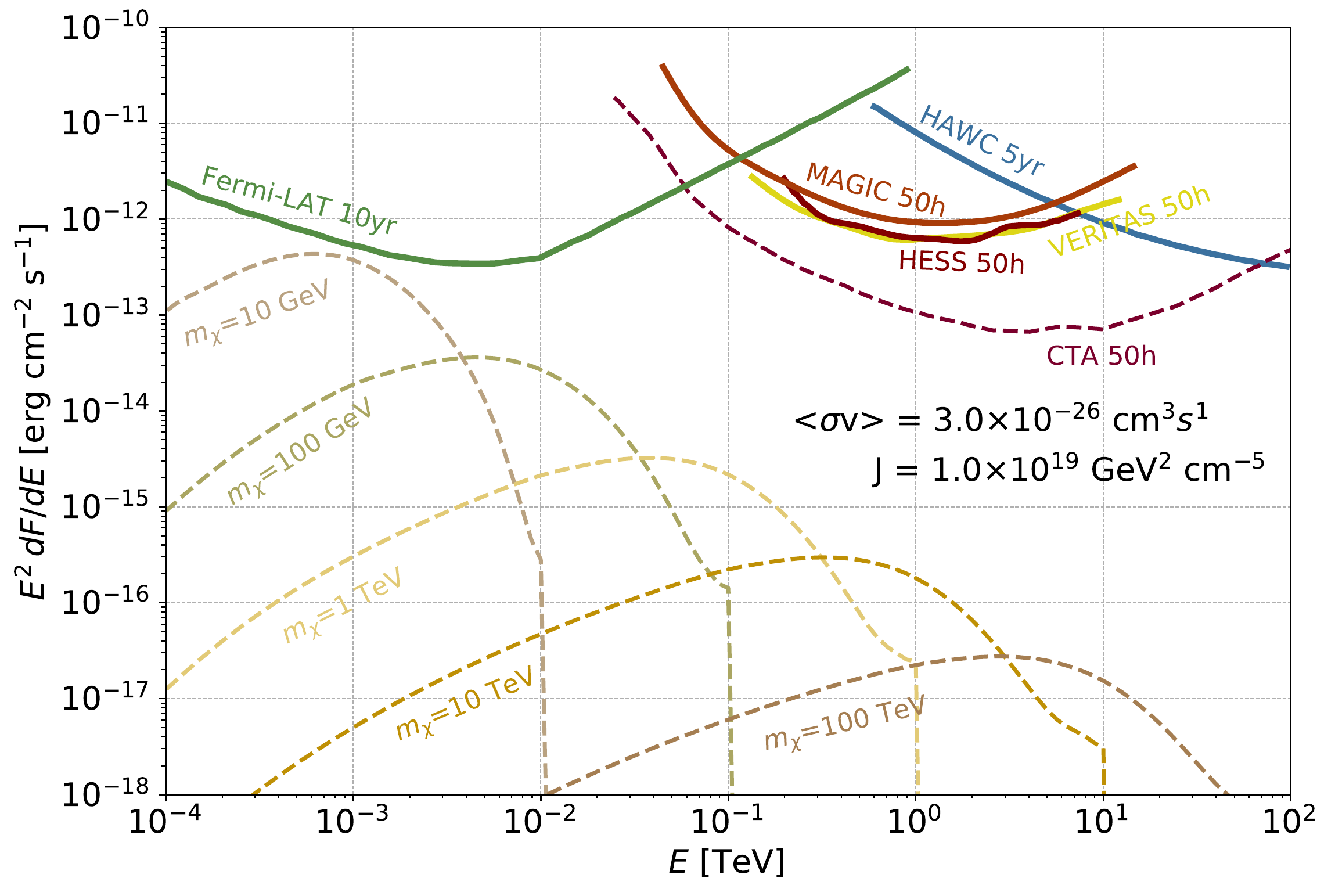}
\includegraphics[width=7.75 cm]{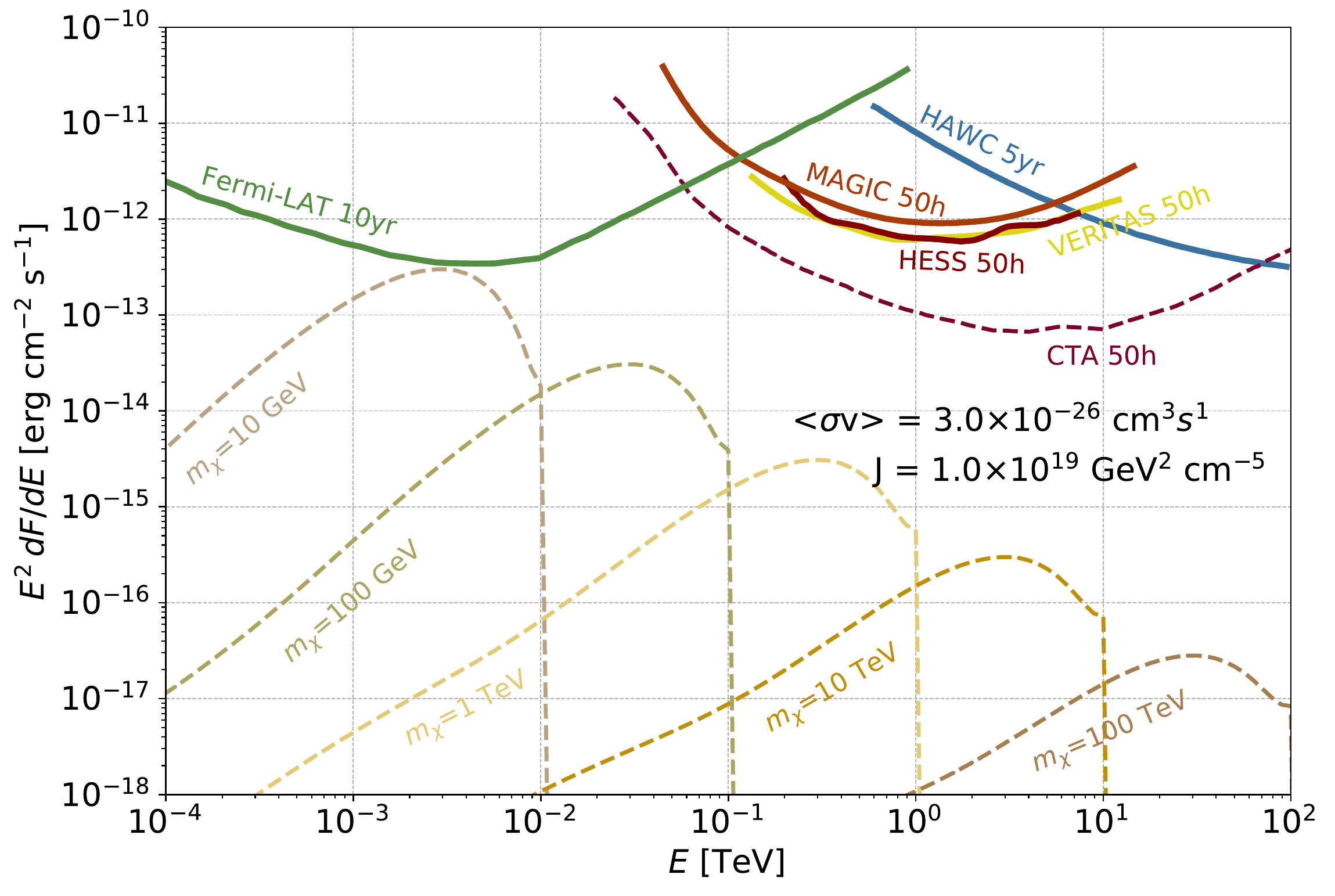}
\caption{Expected gamma-ray spectral energy distribution for WIMPs of
  masses $\mdm = 0.01, 0.1, 1, 10$ and 100 TeV annihilating with
  $\sv = 3\times 10^{-26}$ \svunits\ into $\bb$ (left) and $\tautau$
  (right) pairs in a dSph with associated J-factor 
  $\Jann = 5\times 10^{21}$ \Junits; also shown are the sensitivity curves
  for the instruments considered in this paper. Fermi-LAT sensitivity
  curve~\cite{ref:LATPass8IRF} corresponds to observations of a
  point-like source at Galactic coordinates
  $(l,b)=(120^\circ,45^\circ)$ for 10 years, analyzed using the latest
  (Pass8) data reconstruction tools; HESS~\cite{ref:HESSIRF},
  MAGIC~\cite{ref:MAGICIRF} and VERITAS~\cite{ref:VERITASIRF} curves
  correspond to 50~h of observations of a point-like source at low
  (Zd $\lesssim 30^\circ$) zenith distance; 
  HAWC curve~\cite{ref:HAWCIRF} is for five years of observations of a
  point-like source at a declination of +22$^\circ$N. The flux
  sensitivity for 50~h observations with the future Cherenkov
  Telescope Array~\cite{ref:CTAopt} is shown for comparison.}
\label{fig:sensitivity}
\end{figure}

The expected primary products of the WIMP annihilation and decay
processes are pairs of leptons, quarks or gauge bosons, which would
produce secondary gamma-rays (among other stable products) through
final-state radiation or hadronization+decay chains. It is
straightforward to compute the contribution to $\dNdE$ from the
different annihilation/decay channels, for a given WIMP mass, using
standard Monte Carlo simulation packages such as
PYTHIA~\cite{ref:pythia81}. The spectral energy distribution of the
gamma-ray continuum resulting from these processes peaks between one
and two orders of magnitude below the WIMP mass, depending on the
channel, as shown in Figure~\ref{fig:sensitivity}. The plots show that
Fermi-LAT is the most sensitive instrument for searching for WIMPs up
to a dark matter mass ($\mdm$) of few TeV in the case of $\bb$ channel
and of few 100 GeV for the $\tautau$ channel. Cherenkov telescopes
dominate the search between those masses and $\sim 100$ TeV for $\bb$
and few 10 TeV for $\tautau$, and HAWC for even higher WIMP
masses. Primary gamma rays like, e.g., those from the
$\chi [\chi] \to \gamma \gamma$ or $\chi [\chi] \to \gamma Z$
processes would be [quasi-]monochromatic.  These would constitute the
cleanest possible dark matter signal, given how there is no known
astrophysical process able to produce such gamma-ray spectral lines,
and that backgrounds affecting the measurement could be drastically
reduced using spectral criteria. If detected, a gamma-ray line would
by itself be considered a clear evidence for the presence of dark
matter. However, due to parity conservation, primary gamma rays can
only be produced via loop processes, which significantly reduces their
associated rate $\Gdm$.

It is useful to particularize the line of sight integral in
Equation~(\ref{eq:flux}) for the annihilation and decay~cases:

\begin{itemize}

\item For annihilation, $\Gdm = \frac{1}{k}\ndm\sv$, with $\sv$ the
  average of the product of the WIMP velocity and annihilation cross
  section. The value of $k$ depends on whether WIMPs are Majorana
  ($k=2$, to take into account that an annihilation involves two
  identical particles) or Dirac particles ($k=4$, reflecting
  the fact that particles can only annihilate with their---equally
  abundant---antiparticles). Including this into
  Equation~(\ref{eq:flux}), and writing the WIMP number density $\ndm$
  in terms of its mass and density ($\rho$), we obtain:
\begin{equation}
\frac{d^2\Phi_{\mathrm{ann}}}{d\Omega\, dE} (E,\bp)= \frac{1}{4\pi}\,
\frac{\sv}{k\, \mdm^2}\, \frac{d\Jann} {d\Omega}(\bp) \dNdE(E)\quad ,
\label{eq:diffgammaflux_ann}
\end{equation}
where we have defined the {annihilation differential
  J-factor} as:
\begin{equation}
\dJdpann(\bp)= \intlos dl \rho^2(\bp,l)\quad .
\end{equation}

\item For decay, the rate is given simply by the
  inverse of the dark
  matter decay lifetime, i.e.,  $\Gdm = \taudm^{-1}$, since each WIMP
  particle decays independently
  of each other. Including this into Equation~(\ref{eq:flux}), we get:
\begin{equation}
\frac{d^2\Phi_{\mathrm{dec}}}{d\Omega\, dE} (E,\bp) = \frac{1}{4\pi}\, \frac{1}
{\taudm \mdm}\, \frac{d\Jdec}{d\Omega}(\bp)\, \dNdE(E)\quad ,
\label{eq:diffgammaflux_ann}
\end{equation}
where we have defined the {decay differential J-factor} as:
\begin{equation}
\dJdpdec(\bp) = \intlos dl \rho(\bp,l)\quad .
\end{equation}
\end{itemize}

The J-factor in a region of the sky $\dom$ is given by:
\begin{equation}
J(\dom) = \int_{\dom}\, d\Omega\, \dJdp \quad ,
\end{equation}
both for $\Jann$ and $\Jdec$. It is convenient to define the
{total} J-factor for a given dSph as: 
\begin{equation}
\Jtot \equiv J(\domtot) \quad ,
\label{eq:Jtot}
\end{equation}
with $\domtot$ a region of the sky containing the whole dSph dark matter
halo. The differential J-factor can be written as:
\begin{equation}
\dJdp(\bp) = \Jtot \cdot \djdp(\bp) \quad, 
\end{equation}
where $\djdp$ can be interpreted as the
PDF for the arrival direction of gamma rays produced by dark matter
processes in the dSph halo, since $\int_{\domtot}\, d\Omega\, \djdp =
1$. Using this notation, the differential gamma-ray flux per energy
and solid angle can be written as: 
\begin{equation}
\frac{d^2\Phi}{dE d\Omega}(E,\bp) = a\, \Jtot\, \djdp(\bp)\, \dNdE(E) \quad ,
\end{equation}
(with $a$ being either $a_{\mathrm{ann}} \equiv \frac{1}{4\pi} \frac{\sv}{k\,\mdm^2}$ for annihilation or
$a_{\mathrm{dec}}\equiv \frac{1}{4\pi} \frac{1}{\taudm\mdm}$ for decay). 
The {differential flux per unit energy} is given by:
\begin{equation}
  \frac{d\Phi}{dE}(E) \equiv \int_{\domtot} d\Omega\, \frac{d^2\Phi}{dE
  d\Omega}(E,\bp) = a\, \Jtot\, \dNdE (E) \quad .
\label{eq:dPhidE}
\end{equation}

The distribution of dark matter within the halo, $\rho(\bp,l)$, is
usually estimated by solving the spherical Jeans equation for the
stellar kinematic data~\cite{ref:Strigari2007}. Using this technique,
several authors have produced catalogues of J-factors for the
different known dSphs. In general, the classical dSphs, with
relatively large stellar populations ($O(100-1000)$), have relatively
low associated J-factors (typically between $3\times 10^{17}$ and
$7 \times 10^{18}$~GeV$^2$cm$^{-5}$ within an integrating angle of
$0.5^\circ$), with associated uncertainties also relatively low
(typically below $50\%$), suitable for setting robust limits to dark
matter properties. On the other hand, members of the ultra-faint
population (those discovered by the Sloan Digital Sky Survey or later,
with $O(10-100)$ members stellar populations) can have larger
estimated J-factors (some above $10^{19}$~GeV$^2$cm$^{-5}$) but also
larger uncertainties (some above a factor 10), therefore providing
better prospects for discovery but less robust constraining power. A
detailed review about the expected dark matter content and
distribution of the known dSphs can be found elsewhere in this
volume.

\section{Gamma-Ray Telescopes}
\label{sec:instruments}

For WIMP indirect searches with gamma-rays, the relevant energy range
spans from 100~MeV to 100~TeV (see Figure~\ref{fig:sensitivity}).
Photons of these energies interact in the upper layers of the
atmosphere, making impossible their {direct} detection from the
ground. Several different experimental techniques have been developed
to detect gamma rays, each optimized for a different energy range and
hence for different dark matter masses.

At energies below $\sim$100 GeV, we can efficiently measure gamma rays
before their destructive interaction in the atmosphere by direct
detection with balloon or satellite-borne detectors. Gamma rays
interact within the detector, and convert into $e^+e^-$ pairs, which
are tracked to estimate the direction of the primary particle, and
then stopped by a calorimeter to estimate its energy. This method is
limited by the relatively small achievable collection area,
corresponding essentially to the physical size of the detector. On the
other hand, the technique presents the great advantages of $\sim$100$\%$
duty cycle, large field of view, and that the much more abundant
charged cosmic rays can be easily identified and therefore vetoed,
resulting in virtually background-free gamma-ray
measurements. Currently, the most advanced gamma-ray
telescope using this detection technique is the Fermi-LAT. It consists of a
large-field-of-view (2.4 sr), pair-conversion telescope, sensitive to
gamma rays in the energy range between 20~MeV and about
300~GeV~\cite{ref:Atwood2009}. The latest Fermi-LAT source catalogue
contains about 5000~sources~\cite{ref:LAT8yrCat}, a third of which
remain unassociated. Since its launch in June 2008, the LAT has
primarily operated in survey mode, scanning the whole sky every 3
h. The exposure coverage of this observation mode is fairly
uniform, with variations below 30\% with respect to the average
exposure. Thanks to this full-sky coverage, Fermi-LAT will be able to
perform dark matter searches using its data archive should new dSphs
be discovered in the future.

Above few tens of GeV, gamma-ray fluxes become too low for the
relatively small collection area of Fermi-LAT, and it is advantageous
to measure them {indirectly} through the detection of the
secondary particles and/or the radiation present in the particle
cascade resulting from their interaction in the atmosphere, which
greatly increases the effective collection area.

Cherenkov telescopes measure the Cherenkov radiation emitted by the
electrons and positrons of the cascade (which travel faster than light
in the atmosphere), thus producing an image of such cascade. The
intensity, orientation, and shape of Cherenkov images allow for the
estimation of the energy and arrival direction of the primary
particle, and provide some separation power between gamma rays and
charged cosmic rays. Several nearby telescopes observing the same
gamma-ray source may image the same cascade from different
perspectives, increasing the precision of these measurements. The weak
points of this technique are the small duty cycle (about 10--15$\%$,
since they operate only during night, with no or relatively dim
moonlight and good atmospheric conditions), narrow fields of view of
few degrees diameter at most, and the presence of the irreducible
background produced by charged cosmic rays. Among its advantages, we
find the large collection area, given by the size of the Cherenkov
light pool projected on the plane of the telescope reflector (e.g.,  
$\sim$$10^5$~m$^2$ for 1 TeV gamma ray at low zenith distance). The
resulting flux sensitivity achieved by this technique reaches
currently around $\sim 1\%$ of the Crab nebula in 25~h of
observations. There are three main running Cherenkov observatories
exploiting this detection technique: H.E.S.S, MAGIC and VERITAS.
H.E.S.S is composed of four 12-m diameter telescopes operating since
2004, surrounding one 28-m diameter telescope since 2012, located in
the Khomas Highland (Namibia). The energy threshold is 30~GeV and the
field of view has a diameter of 5$^\circ$. MAGIC is composed of two
17-m diameter telescopes, located at the Observatorio Roque de los
Muchachos at La Palma, Canary Islands (Spain), in operation since 2004
in single-telescope mode and 2009 in two-telescope mode. MAGIC energy
threshold is 30 GeV and the FoV is 3.5$^\circ$ diameter. Finally,
VERITAS is composed of four 12-m diameter Cherenkov telescopes,
located at the Fred Lawrence Whipple Observatory, Arizona (USA),
operating since 2007. VERITAS has an energy threshold of 85 GeV and a
FoV of 3.5$^\circ$ diameter.

Finally, water Cherenkov particle detectors measure the charged
particles present in the cascades initiated by the primary gamma rays
when interacting in the atmosphere. The amount of detected particles
and their spatial distribution allow to measure the energy of the
primary and to discriminate between gamma rays and cosmic rays,
whereas the difference of detection time at different detectors allows
to estimate the arrival direction. This technique is sensitive to
gamma rays and cosmic rays between few hundred GeV and 100 TeV. It has
the advantages of $100\%$ duty cycle, plus large effective area and
field of view, but a limited separation power between gamma rays and
cosmic rays. The~currently most advanced water Cherenkov gamma-ray
detector is HAWC, composed of 300 water Cherenkov detectors located at
an altitude of 4100 m at the Sierra Negra volcano, near Puebla
(Mexico), covering 22,000~m$^2$. It is sensitive to gamma rays between
500 GeV and 100 TeV, with a field of view of 15\% of the sky, and
daily coverage of 8.4~sr, or 67\% of the sky (a region where dark
matter searches using the HAWC data archive will be possible should
new dSphs will be discovered in the future). Partial HAWC operations
started in 2013, and the full detector was completed in March 2015.

\section{Statistical Data Analysis}
\label{sec:analysis}

Advanced searches for dark matter annihilation or decay in dSphs with
gamma rays rely on the distinct spatial and spectral features of the
expected signals. We expect dark matter signal to be distributed
morphologically according to $\djdp$, and spectrally according to
$\dNdE$, and those PDFs are in general clearly distinguishable from
those expected for background processes.

Regarding the use of the morphological information, the spatial
coincidence of the signal with the position of the dSph would provide
strong discrimination power, because we do not expect that gamma rays
can be produced at dSphs by any conventional astrophysical
process. However, using the information of the {morphology} of
the gamma-ray emission around the position of the dSph is more
delicate, because such morphology is in general subject to relatively
large uncertainties, and assuming an incorrect shape may bias the
result of the search. In addition, the expected size of the dark
matter halo is, for many of the known dSphs and for the considered
gamma-ray instruments, consistent with point like sources, or at most
slightly extended, which means that we can obtain no or little
signal/background discrimination power from the use of the
morphological information. All this is particularly true for dark
matter {annihilation}, for which, due to the $\rho^2$ dependence
of $\djdp$, the expected signal is more compact and more affected by
uncertainties on the details of the dark matter distribution within
the halo. When looking for dark matter {decay} signal, on the
other hand, such dependence is linear with $\rho$, which leads to less
peaked and less uncertain morphologies.

The use of spectral information would be key for univocally
attributing a dark matter origin to a detected gamma-ray signal,
because in general, the features present in the spectra predicted for
dark matter annihilation or decay cannot be produced by other
conventional astrophysical processes. For instance, in the most
extreme/luckiest case, the detection of gamma-ray spectral lines would
be considered as unambiguous prove for the observation of dark matter
annihilation or decay. Other processes, like creation of Standard
Model particle pairs also produce distinct spectral features providing
high discrimination power over backgrounds, such as the existence of
sharp kinematic spectral cutoffs (see Figure~\ref{fig:sensitivity}).
These considerations are general for all dark matter searches,
independently of whether they are performed on dSphs or elsewhere.
Searches in dSphs have the additional advantage that dark matter
signals are, in principle, {universal}, any potential detection
from a given dSph could be confirmed by looking for the same spectral
features in the emission from other dSphs. Contrary to the case of
$\djdp$, uncertainties in $\dNdE$ can be considered negligible for a
given annihilation/decay channel. This is the main reason why
gamma-ray instruments utilize the spectral information not only for
reinforcing the credibility of an eventual future detection, but also
to increase the sensitivity of the search and therefore provide more
constraining bounds to the dark matter nature in case of no detection.

Current dark matter searches using gamma rays are based on different
implementations of the likelihood-ratio test~\cite{ref:pdg2018}, which
we use to quantify the compatibility of the measured data ($\data$)
with different hypotheses, in particular with the {null}
hypothesis (i.e.,  that no dark matter signal is present in $\data$),
through the associated p-value. Finding a sufficiently low p-value (by
convention in the field $p < 3\cdot 10^{-7}$) for the observed data
$\data$ under the null hypothesis assumption is usually referred to as
{detecting} dark matter. In case of a positive detection, we can
use the likelihood function to measure the dark matter physical
parameters such as its mass, annihilation cross section, decay
lifetime, and branching ratio to the different decay/annihilation
channels (collectively represented here by the vector $\bm{\aaa}$).
Conversely, if the null hypothesis cannot be excluded, we can use the
likelihood function to set limits to the parameters $\bm{\aaa}$.

The likelihood function can be written in the following general form: 
\begin{equation} 
\lkl (\bm{\aaa}; \bm{\nu} | \data) \quad , \label{eq:lkl}
\end{equation} 
where, apart from its dependence on $\bm{\aaa}$ and $\data$, we have
made explicit that $\lkl$ can also depend on other, so-called,
{nuisance} parameters ($\bm{\nu}$), for which we only know their
likelihood function (normally constrained using dedicated datasets).
In general, nuisance parameters represent quantities used in the
computation of $\bm{\aaa}$ and that are affected by some uncertainty,
either of statistical or systematic nature, or~both. Prototypical
examples of nuisance parameters are the number of background events of
certain estimated energy and arrival direction present in the signal
region, or $\Jtot$. One standard technique to eliminate the nuisance
parameters when making statements about $\bm{\aaa}$ is using the
{profile likelihood ratio~test}:
 \begin{equation}
  \lp(\bm{\aaa}\, |\, \data) = \frac{\lkl(\bm{\aaa}; \hat{\hatnu}\,
  |\, \data)}{\lkl(\hat{\bm{\aaa}}; \hatnu\,  |\, \data)}\quad ,
\label{eq:profile}
\end{equation}
where $\hat{\bm{\aaa}}$ and $\hatnu$ are the values maximizing $\lkl$,
and $\hat{\hatnu}$ the value that maximizes $\lkl$ for a given
$\bm{\aaa}$.  According to Wilks’ theorem $-2\ln\lp(\bm{\aaa})$ is
distributed, when $\bm{\aaa}$ are the true values, as a $\chi^2$
distribution with number of degrees of freedom equal to the number of
components of $\bm{\aaa}$, independent of the value of $\bm{\nu}$. It~is an extended practice in indirect dark matter searches with gamma
rays to decrease the n-dimensional vector $\bm{\aaa}$ of free
parameters to a one-dimensional quantity $\aaa$, by considering that
gamma-ray production is dominated either by annihilation ($\aaa =
\sv$, i.e.,  the velocity-averaged annihilation cross section) or by
decay ($\aaa = \taudm^{-1}$, i.e.,  the decay rate), and scanning over
values of the dark matter particle mass ($\mdm$) and pure
annihilation/decay channels (i.e.,  considering at each iteration 100\%
branching ratio to one of the possible SM particle pairs). For each
scanned combination, Equation~(\ref{eq:lkl}) reduces to a likelihood
function of just one purely free (i.e.,  non-nuisance) parameter. In
such a case, for instance, 1-sided 95\% confidence level \textbf{upper
  limits} to $\aaa$ are taken as $\aaanc=\aaadsu$, with $\aaadsu$
found by solving the equation $-2\ln \lp(\aaadsu) = 2.71$. 

The data $\data$ can refer to $\NdSph$ different dSphs, in which case
it is convenient to write the joint likelihood function as:
\begin{equation}
\lkl (\aaa; \bm{\nu} | \data) = \prod_{l=1}^{\NdSph}
\lklgamma(\aaa  \Jtot_l;
\bm{\mu}_l | \data_{\gamma_l}) \cdot \lklJ(\Jtot_l | {\data_{J_l}})
\quad ,
\label{eq:lklwithJ}
\end{equation}
where we have factorized the joint likelihood into the partial
likelihood functions corresponding to each dwarf, and those
subsequently into the parts corresponding to the gamma-ray
observations ($\lklgamma$) and J-factor measurement ($\lklJ$),
respectively; $\Jtot_l$ is the total J-factor (see
Equation~(\ref{eq:Jtot})) of the $l$-th considered dSph, which, as we
have made explicit, is a nuisance parameter degenerated with $\aaa$ in
$\lklgamma$; $\bm{\mu}_l$ represents the additional nuisance
parameters different from $\Jtot_l$ affecting the analysis of the
$l$-th dSph; $\data_{\gamma_l}$ represents the gamma-ray data of the
$l$-th dSph, whereas $\data_{J_l}$ refers to the data constraining~$\Jtot_l$.

For each dSph, we may have $\Nmeas$ independent measurements, each
performed under different experimental conditions, by the same or
different instruments. That is, we can factorize the $\lklgamma$ term
as:
 \begin{equation}
\lklgamma(\aaa  \Jtot; \bm{\mu}| \data_\gamma) = 
\prod_{k=1}^{\Nmeas} \lklgammak(\aaa  \Jtot; \bm{\mu}_k| \data_{\gamma,k})
\quad ,
\label{eq:multiinstrumentlkl}
\end{equation}
where we have omitted the index $l$ referring to the dSph for the sake
of clarity, and with $\bm{\mu}_k$ and $\data_{\gamma,k}$ representing the
nuisance parameters and data, respectively, referred to the $k$-th
measurement.

For each observation of a given dSph under certain experimental
conditions, $\lklgammak$ often consists of the product of
$\NEbins \times \NPbins$ Poissonian terms ($P$) for the observed
number of gamma-ray candidate events ($\Nij$) in the $i$-th bin of
reconstructed energy and $j$-th bin of reconstructed arrival
direction, times the likelihood term for the $\bm{\mu}$ nuisance
parameters ($\lklmu$), with $\NEbins$ the
number of bins of reconstructed energy and $\NPbins$ the number of
bins of reconstructed arrival direction, i.e.: 
\begin{equation}
  \lklgammak(\aaa  \Jtot; \bm{\mu}| \data_\gamma) = 
  \prod_{i=1}^{\NEbins} \prod_{j=1}^{\NPbins}  P\left(s_{ij}(\aaa \Jtot;\bm{\mu}) +
  b_{ij}(\bm{\mu}) | \Nij\right) \cdot \lklmu(\bm{\mu}|\data_\mu)
\quad ,
\label{eq:binnedLkl}
\end{equation}
where the indexes $l$ and $k$ referring to the dSph and the
measurement have been removed for the sake of a clear notation. The
parameter of the Poissonian term is $s_{ij}+b_{ij}$, where $s_{ij}$ is
the expected number of signal events in the $i$-th bin in energy and
the $j$-th bin in arrival direction, computable using $\aaa \Jtot$ as
we will see below; and $b_{ij}$ the corresponding contribution from
background processes. $\data_\mu$ represents the data used to
constrain the values of the nuisance parameters $\bm{\mu}$. We have
made explicit that the uncertainties associated to $\bm{\mu}$ can in
principle affect both the computation of the signal and background
contributions. For instance, uncertainties in the overall energy scale
affect the computation of $s_{ij}$, whereas uncertainties in the
background modeling affect the computation of $b_{ij}$. However,
uncertainties affecting $s_{ij}$ are usually considered to be largely
dominated by the uncertainty in the J-factor and the dependence of
$s_{ij}$ on $\bm{\mu}$ therefore ignored. Thus, $s_{ij}$,
is given by:
\begin{equation}
  s_{ij}(\aaa \Jtot) = \int_{\Delta E'_i} dE' \int_{\Delta\bp'_j} d\Omega'
  \int_0^\infty dE \int_{\domtot} d\Omega \int_0^{\Tobs}\, dt\,
 \frac{d^2\Phi(\aaa \Jtot)}{dE\,d\Omega}\, \IRF(E',\bp'|E,\bp,t) \quad ,
\label{eq:sij}
\end{equation}
where $E'$, $\bp'$, $E$ and $\bp$ are the estimated and true energies
and arrival directions, respectively; $d\Omega'$ and $d\Omega$
infinitesimal solid angles containing $\bp'$ and $\bp$, respectively;
$\Tobs$ the total observation time; $t$ the time along the
observations; and $\IRF$ the instrument response function, i.e.\
$\IRF(E',\bp'|E,\bp,t) \, dE'\, d\Omega'$ is the effective collection
area of the detector times the probability for a gamma ray with true
energy $E$ and direction $\bp$ to be assigned an estimated energy in
the interval $[E',E'+dE']$ and $\bp'$ in the solid angle $d\Omega'$
(see more details below), at the time $t$ during the observations. The
integrals over $E$ and $\bp$ perform the convolution of the gamma-ray
spectrum with the instrumental response, whereas those over $E'$ and
$\bp'$ compute the events observed within the $i$-th energy bin
($\Delta E'_i$) and the $j$-th arrival direction bin
($\Delta \bp'_j$). It must be noted that, defining several spatial
bins within the source produces relatively minor improvement in
sensitivity to dark matter searches for not significantly extended
sources (i.e.,  those well described by a point-like source, as it is
the case for many dSphs)~\cite{ref:Nievas2016}. For significantly
extended sources, on the other hand, using a too fine spatial
binning makes the obtained result more sensitive to the systematic
uncertainties in the dark matter spatial distribution within the dSph
halo. Thus, a realistic optimization of $\NPbins$ based on sensitivity
should balance the gain yielded by the use of more spatial information
and the loss caused by the increase in the systematic uncertainty.

The $\IRF$ can be factorized as the product of the detector
collection area $\Aeff$ ($\Tobs\cdot\Aeff$ is often referred to as
{exposure}), times the PDFs for the energy ($\pdfE$) and incoming
direction ($\pdfP$) estimators, i.e.:
\begin{equation}
\IRF(E',\bp'|E,\bp,t) = \Aeff(E,\bp,t)\cdot \pdfE(E'|E,t)\cdot
\pdfP(\bp'|E,\bp,t)\quad ,
\end{equation}
where, following the common practice, the (small) dependence of
$\pdfE$ with $\bp$ has been neglected. $\pdfP$ is often referred to as
the {point spread function} (PSF).

Finally, the likelihood for the total J-factor is usually written as:
\begin{equation}
  \lklJ(\Jtot\, |\, \Jobs,\sigma_J) = \frac{1}{\ln(10) \Jobs \sqrt{2\pi}\sigma_J}
\,
  e^{-\left(\log_{10}(\Jtot)-\log_{10}(\Jobs)\right)^2/2\sigma_J^2}\quad;
\label{eq:JfactorPDF}
\end{equation}
with $\log_{10}\Jobs$ and $\sigma_J$ the mean and standard deviation
of the fit of a log-normal function to the posterior distribution of
the {total} J-factor~\cite{ref:Fermi2015}. Therefore, including
$\lkl_J$ in the joint likelihood is a way to incorporate the
{statistical} uncertainty of $\Jtot$ in the estimation of $\aaa$.
It is worth noting that, because $\aaa$ and $\Jtot$ are degenerate, in
order to perform the profile of $\lkl$ with respect to $\Jtot$ it is
sufficient to compute $\lklgamma$ vs $\aaa$ for a fixed value of
$\Jtot$, which facilitates significantly the computational needs of
the profiling operation (see details in footnote 12 of
reference~\cite{ref:MAGICLAT2016}). Including $\Jobs$ {systematic}
uncertainties is much more complex, since they depend mainly on our
choice of the dark matter halo density profile function (e.g.,~NFW~\cite{ref:NFW1997}, Einasto~\cite{ref:Essig2010}, etc.), and there
is no obvious way of assigning a PDF to that choice. Because of this,
the impact of that uncertainty in the bounds in $\aaa$ are usually
roughly quantified by performing the likelihood analysis several
times, each assuming different fitting functions, and comparing the
results obtained for each of them. 

The PDF of the test statistic $-2\ln\lp$ for the no-dark matter null
hypothesis, i.e.,  when the true value of $\aaa$ is given by
$\aaatrue=0$, is needed for evaluating the significance of a possible
signal detection. Computing upper limits to $\aaa$, on the other hand,
consists in finding the value of $\aaatrue$ for which the integral of
the PDF above $\hataaa$ corresponds to the required confidence
level. Estimating the PDF for $-2\ln\lp$ with fast simulations is
feasible (from a computational-demand standpoint) when the involved
p-values are high enough so that they can be evaluated with a
relatively low number of simulated datasets. In practice, however,
results for dark matter searches using gamma-rays are generally
computed assuming Wilks' theorem validity, and that $-2\ln\lp$ is
distributed as a $\chi^2$. The~adoption of Wilks' theorem by all the
experiments allows at least a direct comparison among their results.
One~should keep in mind, however, that the described statistical
framework is also usually affected by the non-fulfillment of the
conditions of validity of Wilks' theorem, at least because of two
different reasons. First, because $\aaa$ is normally restricted to the
physical region (i.e.,  to non-negative values), which produces
over-coverage (i.e.,  the computed confidence interval contains the true
value more often than the quoted confidence level) for negative
background fluctuations, i.e.,   when the likelihood absolute maximum
lies at the border of the physical region. This can be avoided by
using the correct $-2\ln\lp$ PDF for this
situation~\cite{ref:Chernoff}.
%Note that this problem would not arise should
%confidence intervals partially or totally contained in the
%non-physical region were considered acceptable results (what they are
%from a pure statistical point of view because they fulfill the
%definition of coverage). 
Another way commonly used to partially mitigate this problem is to show
the obtained result (e.g.,  the upper limit to $\aaa$) in comparison to
its PDF for the no-dark matter ($\aaatrue=0$) hypothesis. Such PDF is
estimated using fast simulations and/or pure-background datasets (such
as those obtained by considering randomly selected directions as
potential DM targets), and it is normally characterized by its median
(referred to as the {sensitivity} of the measurement) and the
bounds for some predefined (e.g.,  68\%, 95\%, etc.) symmetric
containment quantiles. By such comparison one can evaluate whether the
obtained result is significantly incompatible with the $\aaatrue=0$
hypothesis. The~second violation of Wilks' theorem validity conditions
affects the computation of confidence intervals (i.e.,  the PDF of the
test $-2\ln\lp$ for $\aaatrue > 0$). In this case, however, because
$\aaa$ and $\Jtot$ are degenerate in the likelihood function, the
log-normal shape of the likelihood term $\lklJ$ (see
Equation~(\ref{eq:JfactorPDF})) results in the loss of Gaussianity of
the likelihood for $\aaa$ required by the Wilks' theorem.

As we will see in the next Section, the most common simplifications
adopted in gamma-ray data analyses consist in ignoring the statistical
and/or systematic uncertainties in $\Jtot$ or in the background
contribution to the signal region. Omitting these relevant
uncertainties in general improves artificially the reported
sensitivity and bounds obtained by the analysis, which must be
taken into account when comparing results obtained under different
assumptions.

\section{Results}
\label{sec:results}

None of the different gamma-ray telescopes has obtained a significant
detection in their search for dark matter signals from dSphs.
Therefore, they provide results in the form of upper limits to the
annihilation cross section or lower limits to the decay lifetime. In
this section, I summarize the results obtained by the different
considered instruments. In addition, I highlight and motivate the main
analysis choices adopted by the different experiments as well as the
differences with respect to the general framework described in
Section~\ref{sec:analysis}, also summarized in
Table~\ref{tab:summary}.

\subsection{Fermi-LAT}
\label{sec:LAT}

The Fermi-LAT data are publicly available and several authors outside
the Fermi-LAT Collaboration have searched for DM annihilation signals
in dSphs (e.g., references\ \cite{ref:Mazziotta2012, ref:Drlica2015,
  ref:Baushev2012, ref:Cotta2012, ref:Scott2010,
  ref:Hoof:2018hyn,Li:2018rqo,Zhao:2016xie,Baring:2015sza}). The
Fermi-LAT Collaboration has carried out several searches for dark
matter signals from dSphs, corresponding respectively to 11 months
observations of 14 dSphs~\cite{ref:Fermi2010}, 24 months of
observations of 10 dSphs~\cite{ref:Fermi2011}, 4~years~\cite{ref:Fermi2014} and 6 years~\cite{ref:Fermi2015} of data of
25 dSphs. Here we concentrate on this latter work.

In their 6-year-data search, the Fermi-LAT Collaboration applied their
most developed data (re-)analysis, known as {Pass 8}. They
subsequently searched for gamma-ray signals individually in 25~dSphs
(including the classical and the ultra-faint ones discovered by the
Sloan Digital Sky Survey~\cite{ref:ultra-faint}), and combined the 15
targets with better determined dark matter content. The~dark matter
distribution in each dSph was parameterized using the
Navarro-Frenk-White (NFW) profile~\cite{ref:NFW1997}, constrained
using the prescription by Martinez (2015)~\cite{ref:Martinez2015}. The
$\dNdE$ average spectra for the different considered channels, on the
other hand, were obtained from the PHYTIA-based~\cite{ref:pythia81}
DMFIT package~\cite{ref:dmfit}.

% start a new page
\newpage
% change it to landscape
\paperwidth=\pdfpageheight
\paperheight=\pdfpagewidth
\pdfpageheight=\paperheight
\pdfpagewidth=\paperwidth
\newgeometry{layoutwidth=297mm,layoutheight=210 mm, left=2.7cm,right=2.7cm,top=1.8cm,bottom=1.5cm, includehead,includefoot}
\fancyheadoffset[LO,RE]{0cm}
\fancyheadoffset[RO,LE]{0cm}
%%%%%%%%%%%%%%%%%%%%%%%%%%%%%%%%%%%%%%%%%%%%%%%

%\begin{landscape}
%\begin{table}
\begin{table}[H]
  \caption{Summary of dark matter searches with
    gamma-ray instruments. From left to right, columns show: 
    Bibliographic reference; Instrument; Targets; Investigated decay and/or
    annihilation channels; $dN/dE$ source; J-factor source; whether the following aspects have been included in the
    analysis: J-factor uncertainty, 
    morphology of the source, restriction of $\aaa$ to physical
    region, statistical and
    systematic background uncertainties, determination of the true
    $-2\ln\lp$ PDF; other relevant differences of
    the analyses with respect to the general framework (1: In Equation~\ref{eq:fermiterms}, assuming
    $d\phi/dE \propto E^{-2}$ and energy resolution and bias
    disregarded; 2: In Equation~(\ref{eq:sij}), $\Aeff$ dependence on
    $\bp$ disregarded; 3: In Equation~(\ref{eq:sij}), effect of
    angular resolution disregarded (i.e., $\pdfP \to
    \delta(\bp-\bp')$); 4: In Equation~\ref{eq:fsEtheta}, $f_s$
    assumed radially symmetric with respect to the center of the
    dSph). See main text
    for more details.}
  \centering
  \tablesize{\footnotesize} 
\begin{tabular}{cccccccccccccc}
\toprule
  &                                     &                            &                               &                           &                             &                            &                      &                                                      &\multicolumn{2}{c}{$\bm{\Delta}$ \textbf{bkg} } &                      &                          \\
  \textbf{reference}	& \textbf{Instrument}	& \textbf{dSphs} & \textbf{Channels} &  $\bm{dN/dE$} & \textbf{J-Factor} &  $\bm{\Delta J}$ & \textbf{Ext}  & $\bm{\aaa \geq 0}$ &  \textbf{sta}        &   \textbf{sys}            & \textbf{PDF} &  \textbf{Other} \\
  \midrule
  \cite{ref:Fermi2015} 	& Fermi-LAT 	& \parbox[t]{4.3cm}{Bo\"otes~1, Canes Venatici II, Carina,  Coma Berenices, Draco, Fornax, Hercules, Leo II, Leo IV,  Sculptor, Segue 1, Sextans,  Ursa Major II, Ursa Minor, Willman 1} & \parbox[t]{3.7cm}{ $\bb$, $\tautau$, $\ee$, $\uu$, $\mumu$, $\WW$ [annihilation] }  & PYTHIA 8.1 \cite{ref:pythia81}  & \parbox[t]{2cm}{Following Martinez \cite{ref:Martinez2015}, assuming NFW \cite{ref:NFW1997}} & \checkmark   & \checkmark  &  \checkmark &  $\times$  &  $\times$  &  $\times$ & $^1$ \\
  \midrule
  \cite{ref:Abramowski2014} 	& H.E.S.S	& \parbox[t]{4.3cm}{Carina, Coma Berenices, Fornax, Sagittarius, Sculptor} & \parbox[t]{3.7cm}{ $\bb$, $\tautau$, $\ee\ee$, $\mumu$, $\mumu\mumu$, $\WW+ZZ$ [annihilation] }  & Cembranos et al.~\cite{ref:Cembranos2011}  & \parbox[t]{2cm}{Martinez \cite{ref:Martinez2015} assuming NFW \cite{ref:NFW1997} and Burkert \cite{ref:Burkert1995}} & \checkmark   &  $\times$   & \checkmark  & \checkmark &  $\times$  &  $\times$  & $^{2,3}$\\
  \midrule
  \cite{ref:Abdalla2018} 	& H.E.S.S	& \parbox[t]{4.3cm}{Carina, Coma Berenices, Fornax, Sagittarius, Sculptor} & \parbox[t]{3.7cm}{$\gamma\gamma$ [annihilation]}  & trivial  & \parbox[t]{2cm}{Geringer-Sameth et al. \cite{ref:Geringer2014}} & \checkmark   & \checkmark  &  \checkmark & \checkmark &  $\times$  &  $\times$  & $^{2,3}$\\
  \midrule
  \cite{ref:MAGICSegue1} 	& MAGIC	& \parbox[t]{4.3cm}{Segue 1} & \parbox[t]{3.7cm}{ $\bb$, $\ttbar$, $\mumu$, $\tautau$, $\WW$, $ZZ$, [annihilation and decay], $\gamma\gamma$, $Z\gamma$,  $\mumu\gamma$, $\tautau\gamma$ [annihilation], $\gamma\nu$, $\gamma\gamma\gamma\gamma$ [decay] }  & Cembranos et al.~\cite{ref:Cembranos2011}  & \parbox[t]{2cm}{Essig et al. \cite{ref:Essig2010}} &  $\times$   & \checkmark  &  $\times$  &  $\times$   &  $\times$  &  $\times$  & -- \\
  \midrule
  \cite{ref:MAGICUrsaMajorII} 	& MAGIC	& \parbox[t]{4.3cm}{Ursa Major II} & \parbox[t]{3.7cm}{ $\bb$, $\mumu$, $\tautau$, $\WW$, [annihilation]}  & PPC4DMID \cite{ref:Cirelli2011}  & \parbox[t]{2cm}{Geringer-Sameth et al. \cite{ref:Geringer2014}} & \checkmark  & \checkmark & \checkmark & \checkmark & \checkmark  & $\times$  & -- \\
% \midrule
%  \cite{ref:MAGICLAT2016} & \parbox[t]{1.cm}{MAGIC/ Fermi-LAT}	& \parbox[t]{4.3cm}{Bo\"otes~1, Canes Venatici II, Carina,  Coma Berenices, Draco, Fornax, Hercules, Leo II, Leo IV,  Sculptor, Segue 1, Sextans,  Ursa Major II, Ursa Minor, Willman 1} & \parbox[t]{3.7cm}{ $\bb$, $\mumu$, $\tautau$, $\WW$ [annihilation] } & PYTHIA 8.2\cite{ref:pythia82}  & \parbox[t]{2cm}{Martinez \cite{ref:Martinez2015} assuming NFW \cite{ref:NFW1997}} & \checkmark   & \checkmark  & \checkmark & \checkmark/ $\times$  &  $\times$  &  $\times$  & ??\\
  \midrule
  \cite{Archambault2017} & \parbox[t]{1.cm}{VERITAS}	& \parbox[t]{4.3cm}{Bo\"otes~1, Draco, Segue~1, Ursa Minor, Willman~1}  & \parbox[t]{3.7cm}{$\uu$, $\dd$, $\ssbar$, $\bb$, $\ttbar$, $\ee $, $\mumu$, $\tautau$, $\WW$, $ZZ$, $hh$[annihilation] } & PPC4DMID \cite{ref:Cirelli2011}   & \parbox[t]{2cm}{Geringer-Sameth et al. \cite{ref:Geringer2014}} &  $\times$    & \checkmark  & \checkmark &  $\times$  &  $\times$  & \checkmark & $^{2,4}$\\
  \midrule
  \cite{ref:Albert2018} & \parbox[t]{1.cm}{HAWC}	& \parbox[t]{4.3cm}{Bo\"otes~1, Canes Venatici I, Canes Venatici II, Coma Berenices, Draco, Hercules, Leo I, Leo II, Leo IV, Segue 1, Sextans, Triangulum II, Ursa Major I, Ursa Major II, Ursa Minor}  & \parbox[t]{3.7cm}{$\bb$, $\ttbar$, $\mumu$, $\tautau$, $\WW$ [annihilation and decay] } & PYTHIA 8.2~\cite{ref:pythia82}   & \parbox[t]{2cm}{CLUMPY \cite{ref:Clumpy} assuming NFW~\cite{ref:NFW1997}} & $\times$    &  $\times$    & \checkmark & \checkmark &  $\times$  &  $\times$  & -- \\
\bottomrule
\end{tabular}
\label{tab:summary}
\end{table}
%\end{table}
%\end{landscape}

%%%%%%%%%%%%%%%%%%%%%%%%%%%%%%%%%%%%%%%%%%%%%%%
% change everything back
\newpage
\restoregeometry
\paperwidth=\pdfpageheight
\paperheight=\pdfpagewidth
\pdfpageheight=\paperheight
\pdfpagewidth=\paperwidth
\headwidth=\textwidth

Fermi-LAT data statistical analysis follows
Equations~(\ref{eq:lklwithJ}) and (\ref{eq:multiinstrumentlkl}), with
$\NdSph=15$ observed dSphs and $\Nmeas=4$ referring to the four
independent datasets, each containing events with one of the four
possible event direction reconstruction quality level, and hence each
described by different $\IRF$. They consider bins of reconstructed
energy in the range between 500 MeV and 500 GeV and bins of incoming
direction in a region of interest of $10^\circ \times 10^\circ$ around
the position of each dSph. The~dominant background is produced by
gamma rays from nearby sources (whose estimated energy and incoming
direction are consistent, due to the finite angular resolution of the
instrument, with being originated at the dSph dark matter halo), or by
the diffuse gamma-ray component resulting from the interaction of
cosmic rays with the interstellar medium or from unresolved sources of
Galactic and extragalactic origin, depending on the particular dSphs
being considered. The analysis does not explicitly treat the relevant
background parameters $\bm{\mu}$ in Equation~(\ref{eq:binnedLkl}) as
nuisance parameters. Instead, the spectral parameters (e.g.,
normalization, photon index, etc.) of the different background sources
are fixed using the following simplified method. The flux
{normalizations} of the different background components are determined
by means of a maximum-likelihood fit to the spacial and spectral
distributions of the observed events, with the rest of spectral
parameters fixed to the values listed in the updated third LAT source
catalog~\cite{ref:3fgl}. Then, it is checked that the values of the
background normalization factors obtained using this method do not
change significantly by including an extra weak source at the
locations of the dSph, which shows that the background are
well-constrained by this procedure. Studies showed that the effect of
the background uncertainty from this procedure contributed at a few
percent of statistical uncertainty of the signal and are therefore
safe to neglect.

In order to produce a result valid for arbitrary spectral shapes
(i.e.,  arbitrary value of $\mdm$ and of the branching ratios to the
different annihilation/decay channels), the Fermi-LAT Collaboration
computes, for each observed dSph and bin of estimated energy $\Delta
E'_i$, the value of:
\begin{equation}
{\lklgamma}_i(\barPhi_i)=
\prod_{j=1}^{\NPbins}  P(s_{ij}(\barPhi_i) + b_{ij} |
\Nij) 
\label{eq:fermiterms}
\end{equation}
as a function of $\barPhi_i$, that is, the sum over the spatial bins
of the $\lklgamma$ likelihood values within $\Delta E'_i$. In order to
obtain a set of generic ${\lklgamma}_i$ values, they compute
$s_{ij}(\barPhi_i)$ using Equation~(\ref{eq:sij}) assuming a power-law
gamma-ray spectrum ($\frac{d\Phi}{dE} \propto E^{-\Gamma}$) of
spectral index $\Gamma = 2$. The spatial distribution of gammas (which
does not depend on the energy) is considered known and fixed, and
given by the $\dJdp$ curves obtained from the fit to the stellar
kinematics to the different dSphs.  Equation~(\ref{eq:binnedLkl}) can
then be written in terms of the ${\lklgamma}_i$ factors as:
\begin{equation}
\lklgammak(\aaa  \Jtot; \bm{\mu}| \data_\gamma) = 
\prod_{i=1}^{\NEbins} {\lklgamma}_i(\barPhi_i(\aaa \Jtot))
\quad .
\label{eq:fluxLkl}
\end{equation}

The values of ${\lklgamma}_i$ vs $\barPhi_i$ for each of the analyzed
dSphs were computed, tabulated and released by the Fermi-LAT
Collaboration~\cite{ref:publkltables}. This allows any scientist to
compute $\lklgamma$ for the dark matter model of their choice by just
selecting the corresponding values of $\aaa$, $\mdm$, the total
$\dNdE$ and $\Jtot$, and computing the corresponding $\barPhi_i$
values as:
\begin{equation}
\barPhi_i(\aaa \Jtot) = \int_{\Delta E_i}
dE\,  \frac{d\Phi(\aaa \Jtot)}{dE} \quad ,
\end{equation}
with $\frac{d\Phi}{dE}$ given by Equation~(\ref{eq:dPhidE}). We note
that this approach allows to compute bounds on $\aaa \Jtot$ with
associated confidence level known only to a certain (unquantified)
precision that depends on how similar are the investigated spectral
shape and the one assumed when computing the values of ${\lklgamma}_i$
(i.e.,  a power-law with $\Gamma = 2$ in the Fermi-LAT case). In addition, it
should be stressed that such precision depends also on the PDF of the
energy estimator and that, therefore, the range of investigated
spectral shapes for which we can establish bounds within a certain
precision using this technique is different for different instruments.

\begin{figure}[t] 
\centering 
\includegraphics[width=1.0\textwidth]{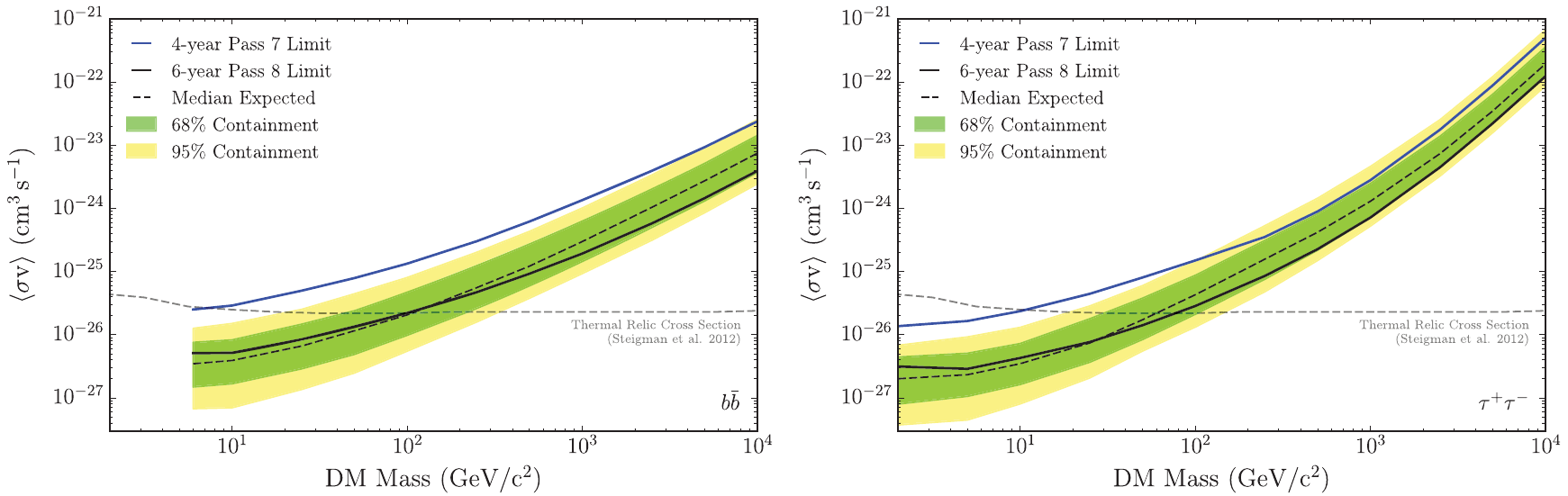}
\caption{The 95\% confidence-level upper limits to $\sv$ for the
  $\chi\chi \to \bb$ (\textbf{left}) and $\chi\chi \to \tautau$
  (\textbf{right}) annihilation channels derived from 6-year
  observations of 15 dSphs with Fermi-LAT. The dashed black line shows
  the median of the distribution of limits obtained from 300 simulated
  realizations of the null hypothesis using LAT observations of
  high-Galactic-latitude empty fields, whereas green and yellow bands
  represent the symmetric 68\% and 95\% quantiles, respectively. The
  dashed gray curve corresponds to the thermal relic
  cross-section~\cite{ref:Steigman2012}.  Reprinted figure with
  permission from reference \cite{ref:Fermi2014}; copyright (2014) by
  the American Physical Society.}
\label{fig:FermiLAT2015limits}
\end{figure}

No significant gamma-ray signal from dSphs was found in the Fermi-LAT
data, either individually in each dSph (the largest deviation from the
null hypothesis is found for Sculptor, with $-2\ln\lp=4.3$), or in the
combined analysis ($-2\ln\lp=1.3$). Some of the obtained exclusion
limits are shown in Figure~\ref{fig:FermiLAT2015limits}. This work
represents the most constraining search for WIMP annihilation signals
for the dark matter particle mass range below $\sim$1 TeV. As shown in
the figure, the limits exclude the thermal relic cross section for
$\mdm < 100$ GeV in the case of annihilation into $\bb$ or $\tautau$
pairs.

These results were combined with MAGIC observations of Segue~1, into
the first coherent search for dark matter using several gamma-ray
instruments~\cite{ref:MAGICLAT2016}. Details about this work are
provided~below.

In a later work, the Fermi-LAT and the Dark Energy Survey (DES)
collaborations also used the data from 6 years of observations to look
for dark matter signals over a sample of 45 stellar systems consistent
with being dSphs~\cite{ref:LATDES2017}. The search was performed
shortly after the discovery of 17 of the considered dSph candidates,
for which no reliable estimate of the dark matter content was 
available at the time. Because of this, all considered candidates were assumed to
be point-like sources, and the J-factors for the non-confirmed dSphs
estimated from a purely empirical scaling relation based on their
heliocentric distance. For four of the examined dSphs, a 2$\sigma$
discrepancy with the null hypothesis was found, which does not
contradict significantly such hypothesis, particularly once the number
of investigated sources, channels and masses is considered. Overall,
the strategy of observing a set of not fully confirmed dSphs
candidates, for which no reliable estimate of the J-factor exists yet
is justified since a solid positive gamma-ray signal from any of the
observed targets would have been considered a strong experimental
evidence of dark matter annihilation or decay. In absence of such
signal, however, the obtained limits are less robust than those from
the 15 confirmed dSphs described above, which remain the reference in
the field for the sub-TeV mass range.

\subsection{Cherenkov Telescopes}

Dark matter searches with Cherenkov telescopes have evolved from
simple event-counting analyses to more complex maximum-likelihood
analyses of optimized sensitivity thanks to the inclusion of the
expected spectral and morphological features of the dark matter
signals~\cite{ref:Aleksic2012}.

In the most basic version of the likelihood function, the nuisance
parameters $\bm{\mu}$ (see Equation~(\ref{eq:binnedLkl})) are the
$b_{ij}$ factors themselves. They are constrained by measurements in
signal-free, background-control (or {Off}) regions with $\tau$
times the exposure of the signal (or {On}) region. A more
complete analysis also includes the treatment of $\tau$ as a nuisance
parameter, given that the latter is normally affected by significant
statistic ($\sigmataustat$) and systematic ($\sigmatausys$)
uncertainties. The statistical uncertainty comes from the fact that
$\tau$ is often estimated from the data themselves (comparing the
events observed in regions adjacent to the On and Off ones). The
systematic uncertainty takes into account the residual differences of
exposure between the Off and On regions, and it is normally assumed to
be of the order of $1\%$ for the current generation of Cherenkov
telescopes~\cite{ref:MAGICIRF}. It can be shown that this systematic
uncertainty is the limiting factor to the sensitivity of the
event-counting analyses for
$\Non \gtrsim \frac{(\tau+1)\tau}{\Delta \tau^2}$, i.e.,   between
$\sim 10^4$ and $\sim 2 \times 10^4$ events for $\tau$ in the typical
range between 1 and 10, and $1\%$ systematic uncertainty in $\tau$ (i.e.,\
$\sigmatausys = 0.01\tau$). Once we reach this number of observed
events in the signal region, increasing the statistics of the dataset
does not longer contribute to improve the sensitivity of the search.

The gamma-ray likelihood function for Cherenkov telescopes can thus be
written as the product of Poisson likelihoods for the On and Off
region times a Gaussian likelihood for $\tau$, i.e.:
\begin{eqnarray}
& & \lklgamma(\aaa \Jtot; \{b_{ij}\}_{i=1,...,\NEbins;
    j=1,...,\NPbins}, \tau \,|\, \{\Nonij, \Noffij\}_{i=1,...,\NEbins;
    j=1,...,\NPbins})  =  \nonumber \\  
 & & \quad \quad \prod_{i=1}^{\NEbins} \prod_{j=1}^{\NPbins}\left[  P\left(s_{ij}(\aaa \Jtot) +
  b_{ij} \,|\, \Nonij\right) \cdot P(\tau b_{ij} \,|\, {\Noffij})
\right] \cdot G(\tau\, |\, \tauobs,\sigmatau)
\quad , 
\label{eq:onoffLkl}
\end{eqnarray}
with $\Nonij$ and $\Noffij$ the number of observed events in the On
and Off regions, respectively, in the $i$-th bin of reconstructed
energy and the $j$-th bin of reconstructed arrival direction; and $G$
an (often neglected) Gaussian PDF with mean the measured value $\tauobs$ and width
$\sigmatau = \sqrt{\sigmataustat^2 + \sigmatausys^2}$.

The considered energy range depends on the instrument (e.g.,  larger
reflectors provide lower thresholds) and the dSph observation
conditions (e.g.,  higher zenith angle observations imply higher
threshold). For the current instruments and observed dSphs, the lowest
energy bin starts between 80~and 800 GeV, whereas the highest one can
reach up to between 10 and 100 TeV.

In the analysis of Cherenkov telescope data, the convolution of
$\frac{d^2\Phi}{dE d\Omega}\cdot \Aeff$ with the PSF function $\pdfP$
needed to compute $s_{ij}$ according to Equation~(\ref{eq:sij}) is
usually performed numerically through the analysis of Monte Carlo
simulated events. We note that Equation~(\ref{eq:sij}) can be written
as:
\begin{equation}
s_{ij} = \int_{\Delta E'_i} dE' 
\int_0^\infty dE\, \int_0^{\Tobs} dt\,
 \frac{d\Phi}{dE}(E)\, \Aeffj(E,\djdp,t)\, f_E(E'|E,t) \quad ,
\label{eq:sij_reduced}
\end{equation}
with $\Aeffj$ the {signal morphology-averaged} effective area
within spatial bin $j$, defined as:
\begin{eqnarray}
  \Aeffj(E,\djdp,t) &  = & \frac{\int_{\Delta\bp'_j} d\Omega' \int_{\domtot} d\Omega\,
    \frac{d^2\Phi}{dE\,d\Omega}(E,\bp)\, \Aeff(E,\bp,t)\,
    \pdfP(\bp'|E,\bp,t)}{\frac{d\Phi}{dE}(E)}\nonumber \\[2mm]
            & = & \int_{\Delta\bp'_j} 
  d\Omega' \int_{\domtot} d\Omega\, \djdp(\bp)\, \Aeff(E,\bp,t)\,
  \pdfP(\bp'|E,\bp,t) \quad .
\label{eq:Aeffj}
\end{eqnarray}
$\Aeffj$ depends on the morphology of the gamma-ray emission
($\djdp$), although not on its intensity, hence not on $\Jtot$.
Therefore, for point-like sources observed with constant IRF at a
given fixed direction, $\Aeffj$ is only a function of the energy. As a
matter of fact, what normally is referred to as {the} effective
area of a given Cherenkov telescope is the value of $\Aeffj(E)$ for a
circular spatial bin centered at the position of a point-like source
(observed at low zenith distance under dark and good weather
conditions), with radius optimized to maximize the signal-to-noise
ratio. In practice, $\Aeffj(E)$ is computed with Monte Carlo
simulations: Using a sample of simulated gamma rays with arrival
directions distributed according to $\djdp$ and trajectories impacting
uniformly in a sufficiently large area ($\atot$) around the telescope
pointing axis, $\Aeffj$ is computed scaling $\atot$ by the ratio
between events detected within spatial bin $j$ and the total number of
generated events. For reasons of economy of computational resources,
$\Aeffj$ is computed in some of the analyses described here
approximating $\djdp$ by a point-like source (i.e.,  by a delta
function), even for the analysis of moderately extended dSphs. This
approximation is less accurate the more extended the source. The bias
introduced in $\Aeffj$ becomes relevant when the source extension is
comparable to or bigger than the region for which $\Aeff$ may be
considered flat.

\subsubsection{H.E.S.S}

The first dark matter searches using observations of dSphs with the
H.E.S.S telescopes were based on an event-counting analysis, with no
attempt to use the expected spectral and morphological signatures in
the search~\cite{ref:Aharonian2008,ref:Abramowski2011a}. H.E.S.S also
performed early searches on non-confirmed dSphs like Canis
Major~\cite{ref:Aharonian2009} or even globular
clusters~\cite{ref:Abramowski2011b}. Their most recent searches use
state-of-the-art analysis techniques like the one described in
Section~\ref{sec:instruments}, and are based on observations of
Sagittarius ($\sim$90~h), Coma Berenices ($\sim$9~h), Fornax
($\sim$6~h), Carina ($\sim$23~h) and Sculptor ($\sim$13~h) dSphs,
where H.E.S.S has searched for both
continuum~\cite{ref:Abramowski2014} and
line-like~\cite{ref:Abdalla2018} dark matter spectra. We will
concentrate in these two latter works.

There are significant differences in the high-level maximum-likelihood
analyses used by H.E.S.S in their searches for continuum and spectral
lines. In the search for continuum spectra the $\dNdE$ was taken from
analytical parameterizations~\cite{ref:Cembranos2011}, generally valid
up to dark matter mass of 8 TeV; and the J-factors were estimated
using the prescription by Martinez (2015)~\cite{ref:Martinez2015},
assuming alternatively cuspy NFW~\cite{ref:NFW1997}, and cored
Burkert~\cite{ref:Burkert1995} profiles. In the search for spectral
lines, on the other hand, the $\dNdE$ is trivial, and the J-factors
were taken from the work by Geringer-Sameth et al.
(2014)~\cite{ref:Geringer2014}, which assumes a Zhao-Hernquist dark
matter density profile~\cite{ref:Zhao1996}. The case of Sagittarius
dSph was treated separately in both works, given that this galaxy is
likely affected by tidal disruption~\cite{ref:Ibata2001}, and
therefore the J-factor calculation is subject to comparatively larger
systematic uncertainties, not included in the likelihood analysis.

There are also slight differences in the likelihood function used by
H.E.S.S in the continuum and spectral line dark matter searches. In
both cases they use the likelihood function of
Equation~(\ref{eq:onoffLkl}) without including the term accounting for
the uncertainty in $\tau$. In the case of continuum spectra, only one
($\NPbins=1$) circular spatial bin centered at the position of each
dSph was considered, whereas for spectral lines $\NPbins=2$ or 3
(depending on the size of the considered dSph), concentric
0.1$^\circ$-width ring-like spatial bins were used. The reason for
this difference must be purely historical (given how the expected and
measured spatial information is essentially common to both searches),
probably in an attempt to increase the sensitivity by including more
information in the likelihood analysis. The~drawback of this approach,
has already been discussed in Section~\ref{sec:analysis}: It can
introduce a bias in $s_{ij}$, with an unquantified effect in the final
sensitivity to $\aaa$. In both analyses, for the computation of
$s_{ij}$ following Equation~(\ref{eq:sij}), the dependence of the
effective area with $\bp$ within the signal region and the effect of
the PSF are ignored. We note that these two simplifications require
opposite conditions: $\Aeff$ can be better approximated by a constant
value for smaller signal regions, i.e.,  smaller dSphs, whereas the
effect of the angular resolution in the distribution of measured
events is smaller for larger dSphs. The effect in the final result of
adopting these two simplifications is not quantified.

\begin{figure}[t]
\centering
\includegraphics[width=8.5 cm]{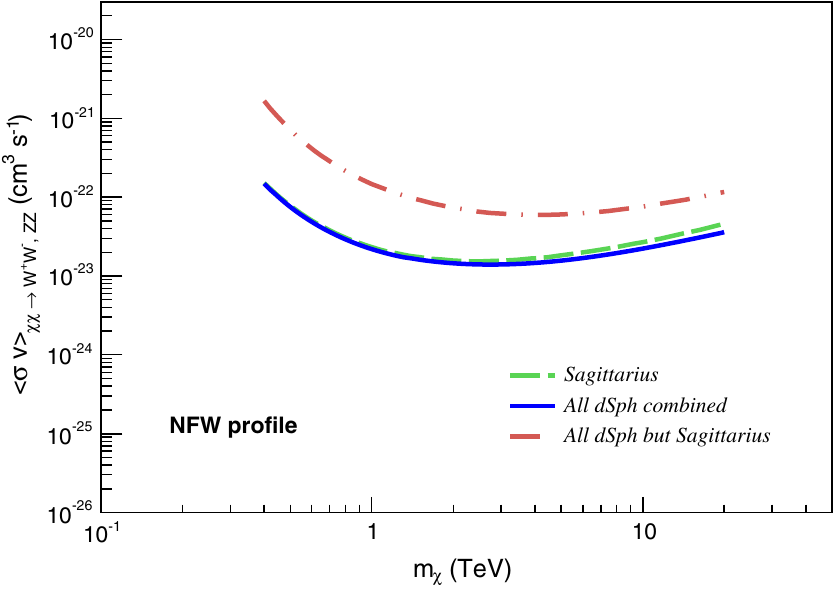}
\includegraphics[width=6.4 cm]{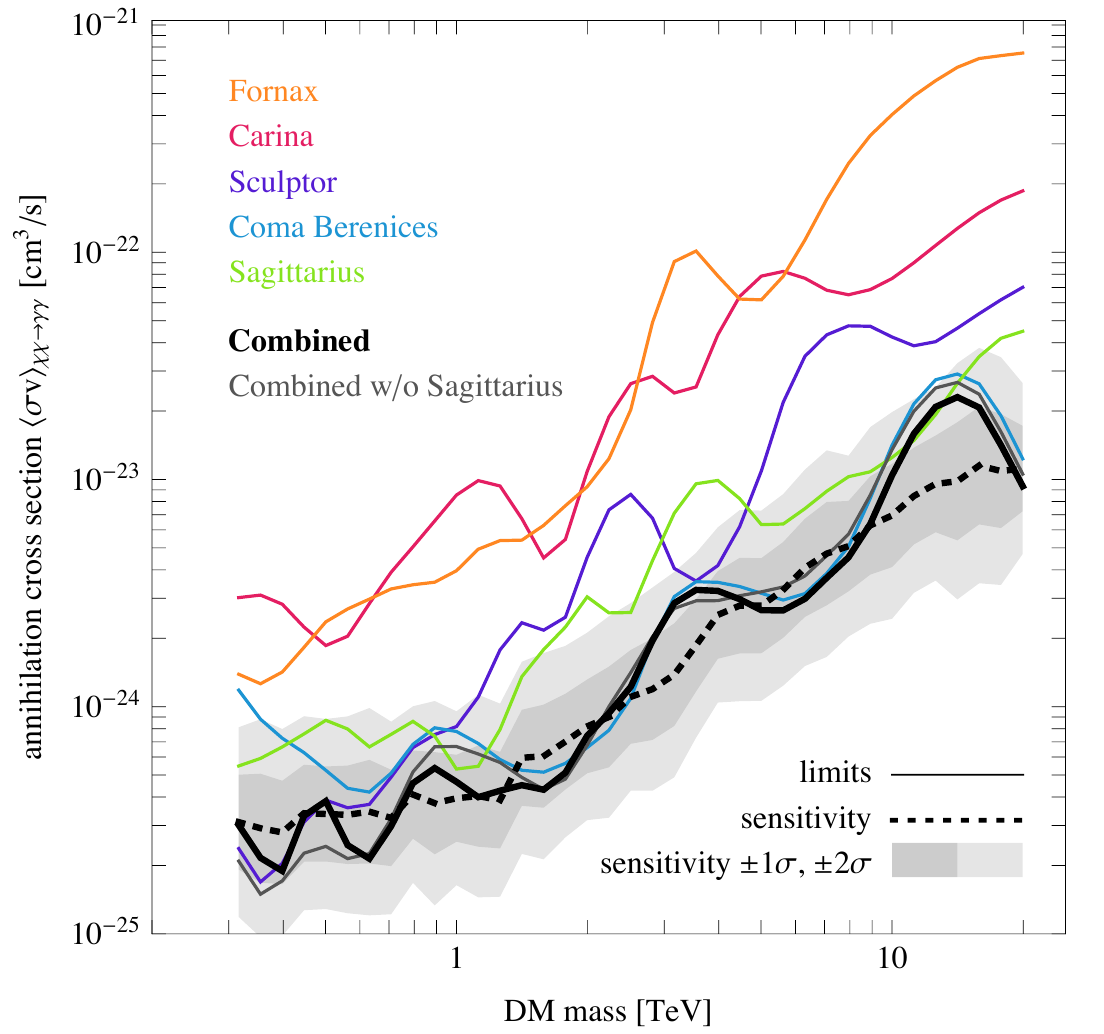}
\caption{The 95\% confidence level upper limits to the cross section
  of dark matter annihilating into a combination of $\WW$ and $ZZ$
  (\textbf{left}, reprinted figure with permission from reference~\cite{ref:Abramowski2014}; copyright (2014) by the American Physical
  Society) and $\gamma\gamma$ pairs (\textbf{right}, reprinted figure
  with permission from reference~\cite{ref:Abdalla2018},
  \textcopyright IOP Publishing Ltd and Sissa Medialab; reproduced by
  permission of IOP Publishing; all rights reserved). Different lines
  show limits from individual dSphs and from their combination with
  and without Sagittarius. For the spectral line search, also the
  median of the distribution of limits obtained for simulated
  realizations of the null hypothesis is shown, together with the
  corresponding 1$\sigma$ and 2$\sigma$ symmetric quantiles.}
\label{fig:HESSlimits}
\end{figure}

H.E.S.S found no significant gamma-ray signal in the observed dSphs,
considered either individually or collectively, for any of the assumed
emission spectra (continuum or spectral line). The~maximum observed
deviations from the null hypothesis are $\sim$$2.6\sigma$ for the
continuum spectra search in Fornax, and  $\sim$$1.2\sigma$ for the
spectral-line search in Sagittarius. The exclusion limits for the
annihilation cross-section for continuum spectra (see
Figure~\ref{fig:HESSlimits}-left) peak at dark matter masses of around
1--2 TeV, depending on the considered channel. Assuming a NFW density
profile, the strongest constraint is provided by Sagittarius dwarf,
with $\svnc \sim 2\times 10^{-23}$~\svunits\ for a combination of
$\WW$ and $ZZ$ annihilation channels. The bounds resulting from the
combination of all the observed dSphs are only marginally better
because Sagittarius has, under the NFW-profile assumption, the largest
by far J-factor among the considered dSph, and because it has
been observed by H.E.S.S for significantly longer time than the rest
of the dSphs. However, given that the value of the J-factor for
Sagittarius is affected by large systematic uncertainties (on account
of the possibility that the system is affected by tidal disruption),
H.E.S.S has also provided constraints obtained from the combination of
all the other dSphs, which results in the limit $\svnc \sim
10^{-22}$~\svunits, for the same annihilation channel. In the case of
the search for spectral lines, limits do not depend significantly on
the inclusion or not of Sagittarius (see
Figure~\ref{fig:HESSlimits}-right), since with the newer approach in
the evaluation of the J-factor used in this work, the limits are
dominated by Coma Berenices results. In the mass range between 400~GeV
and 1~TeV, the obtained limit to the velocity-averaged cross section
is $\svnc \sim 3\times 10^{-25}$~\svunits.

\subsubsection{MAGIC}
\label{sec:MAGIC}
MAGIC performed early dark matter searches using
observations of the dSphs Draco~\cite{ref:MAGICDraco},
Willman~1~\cite{ref:MAGICWillman1} and
Segue~1~\cite{ref:MAGICMonoSegue1}. These searches had a relatively
poor sensitivity, due to the fact that they were based on
one-telescope observations and simple event-counting data analysis.
With the addition of the second telescope, MAGIC dark matter search
strategy was based on deep observations ($\sim$160~h) of the dSph with
the highest J-factor known at that moment, namely
Segue~1~\cite{ref:MAGICSegue1}, and the use for the first time by
Cherenkov telescopes, of advanced maximum-likelihood analysis
techniques. In addition, MAGIC Segue~1 observations were part of the
aforementioned first multi-instrument combined search, together with
data from Fermi-LAT~\cite{ref:MAGICLAT2016}, a work that I will
discuss later in more detail. After that, MAGIC initiated a
diversification of observed targets, starting by $\sim$100~h of
observations of the Ursa Major II dSph~\cite{ref:MAGICUrsaMajorII}.

\begin{figure}[t]
\centering
\includegraphics[width=7.95 cm]{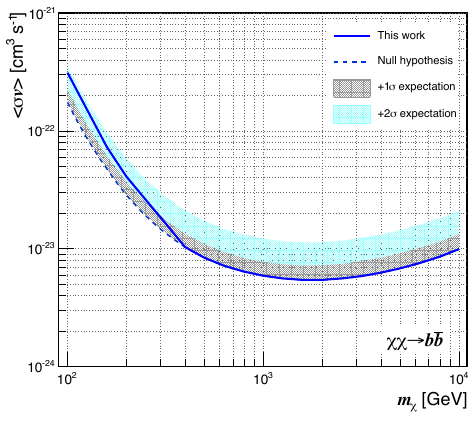}
\includegraphics[width=7.05 cm]{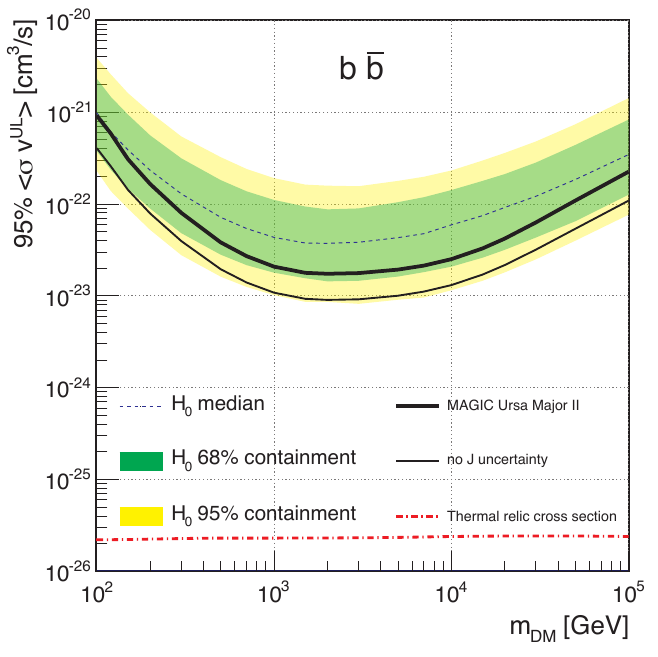}
\caption{The 95\% confidence level upper limits to $\sv$ (solid line)
  for the process $\chi\chi \to \bb$ from observations of the dSphs
  Segue 1 (\textbf{left}, reprinted figure under
  \href{https://creativecommons.org/licenses/by/3.0/}{CC BY license}
  from reference~\cite{ref:MAGICSegue1}) and Ursa Major~II
  (\textbf{right}, reprinted figure with permission from
  reference~\cite{ref:MAGICUrsaMajorII}, \textcopyright IOP Publishing
  Ltd and Sissa Medialab; reproduced by permission of IOP Publishing;
  all rights reserved) with MAGIC; also shown are the median of the
  distribution of limits for the null-hypothesis, and the limits of
  the symmetric 68\% and 95\% quantiles. For Ursa Major~II, both the
  results with and without considering $\Jtot$ statistical uncertainty
  are shown.}
\label{fig:MAGIClimits}
\end{figure}

MAGIC dark matter searches in Segue~1 and Ursa Major~II follow
essentially the same data selection, calibration and processing
procedures, but contain significant differences in several elements of
their high-level analysis. The gamma-ray average spectra per
annihilation reaction ($\dNdE$) were obtained from the
parameterization by Cembranos et al. (2011)~\cite{ref:Cembranos2011}
in the case of Segue~1, and the PPPC 4 DM ID
computation~\cite{ref:Cirelli2011} in the case of Ursa Major~II. The
spectra provided by these two works do not differ significantly for
the considered energy range. The J-factor for Segue~1 was computed
solving the Jeans equation assuming an Einasto density
profile~\cite{ref:Essig2010}, and for Ursa Major~II was taken from
Geringer-Sameth et al. (2014)~\cite{ref:Geringer2014}.

The likelihood function used by MAGIC for dark matter searches has
also evolved over the years. For the observations of Segue~1 they
used, instead of Equation~(\ref{eq:onoffLkl}), the following {unbinned}
likelihood~function: 
\begin{equation} \lklgamma(\aaa \Jtot; b \,|\,
  \{E'_i\}_{i=1,...,\Non}) = P\left(s(\aaa \Jtot) + b \,|\, \Non
  \right) \cdot \prod_{i=1}^{\Non} f_{s+b}(E'_i) \quad ,
  \label{eq:magicunbinnedlkl} 
\end{equation} 
where the uncertainty on $\tau$ is ignored, only one spatial bin is
considered, and the energy-wise product of Poisson terms is
substituted by a global Poisson term for the total number of observed
events ($\Non$), times the joint likelihood for the observed values of
estimated energies. The latter is computed as the product of the PDF
for the reconstructed energy $f_{s+b}(E')$ evaluated at each observed
$E'$, where $f_{s+b} = \frac{1}{s+b}\left(s\, f_s+ b\, f_b\right)$,
with $f_s$ and $f_b$ the PDFs for the reconstructed energies for
signal and background events, and $s$ (the free parameter) and $b$ (a
nuisance parameter) the total expected number of signal and background
events, respectively. $f_s$ is the normalized convolution of the
gamma-ray spectrum with the IRF, i.e.:
\begin{equation}
  f_s (E') = \frac{\Tobs}{s}
  \int_0^\infty dE\, 
 \frac{d\Phi}{dE}\, \avAeff(E)\, f_E(E'|E)
\quad ,
\label{eq:fs}
\end{equation}
with $\avAeff(E)$ computed following Equation~(\ref{eq:Aeffj}). $f_b$,
on the other hand, is modeled using the data from one or several Off
regions. This approach presents the drawback of neglecting the
statistical and systematic uncertainties in the background spectral
shape. In comparison, in the binned version of the likelihood function
(Equation~(\ref{eq:onoffLkl})) the statistical uncertainty is taken into
account by the inclusion of the nuisance parameters $\bij$ and $\tau$.
This unbinned analysis hence typically produces results that are
several tens of percent artificially more constraining than the binned
one. Another important difference of the MAGIC Segue 1 analysis with
respect to the general framework is that it does not include
statistical uncertainties in the J-factor. This was justified by the
fact that the bounds to $\aaa$ scale with $1/\Jtot$, and therefore the
provided results allow the computation of the limits for any other
J-factor value (provided $\djdp$ is kept fixed). This argument is
valid only for single-target observations, but not for results
obtained combining observations from different dSphs with different
$\Jtot$ values and uncertainties. Another main difference between this
analysis and the general framework is the treatment of the cases when
the value $\hataaa$ maximizing the likelihood lies outside the
physical region, i.e.,  $\hataaa<0$. For those cases, the 95\%
confidence limit on $\aaa$ was computed as $\aaanc = \aaadsu-\hataaa$,
with $\aaa$ unrestricted (i.e.,  allowed to take negative values)
during the likelihood maximization process. With this prescription,
the limit obtained for any negative fluctuation in the number of
excess events is equal to the limit for zero excess events (i.e.,  the
sensitivity), at the expense of some over-coverage (i.e.,  the bounds
are conservative).

In the analysis of Ursa Major II data, MAGIC used the general analysis
framework described in Section~\ref{sec:analysis}, with binned
likelihood, statistical uncertainties in the J-factor considered, and
$\aaa$ restricted to positive values. In addition, for the first time
in the analysis of Cherenkov telescope data,
the Off/On exposure ratio $\tau$ in Equation~(\ref{eq:onoffLkl}) was
considered a nuisance parameter, taking into account both its
statistical and systematic ($\sigmatausys = 1.5\%$) uncertainties,
thus providing more realistic results.

MAGIC found no significant gamma-ray signal in the observations of
Segue~1 or Ursa Major~II. This was translated into limits to the dark
matter annihilation cross section (and decay lifetime), assuming
different dark matter induced gamma-ray production mechanisms. Using
Segue~1, MAGIC carried out a systematic search for annihilation and
decay processes, looking for the continuum spectra from production of
$\bb$, $\ttbar$, $\mumu$, $\tautau$, $\WW$ and $ZZ$ pairs, spectral
lines from $\gamma\gamma$ and $\gamma Z$ channels, and other spectral
features such as those produced by virtual internal bremsstrahlung
emission ($XX\gamma$) and gamma-ray ``boxes'' ($\Phi\Phi \to \gamma
\gamma \gamma \gamma$). With Ursa Major~II data, the searches were
limited to annihilation into $\bb$, $\mumu$, $\tautau$ and $\WW$
pairs. Figure~\ref{fig:MAGIClimits} shows the results for annihilation
into $\bb$ pairs obtained from each of the observed dSphs (there is no
MAGIC-only combined result). The obtained limits are in general within
the 68\% containment region expected for the null hypothesis, except
for the low mass range $\mdm \lesssim 300$~GeV in the case of Segue~1, where
they stay nevertheless within the 95\% containment region. 95\%
confidence level upper limits to the annihilation cross-section of
dark matter particles into $\bb$ pairs reach $\svnc \sim 5\times
10^{-24}$~\svunits\ for $\mdm \sim 2$~TeV in the case of Segue~1, and
$\svnc \sim 2\times 10^{-23}$~\svunits\ in the case of Ursa Major~II.
Segue~1 observations were also used to constrain the lifetime of
$\mdm\sim 20$~TeV particles decaying into $\bb$ pairs to be larger
than $\taudmnc \sim 3\times 10^{25}$s.

\subsubsection{VERITAS}

VERITAS has performed dark matter searches using observations of the
dSphs Segue~1 (92~h), Draco (50~h), Ursa Minor (60~h), Bo\"otes~1
(14~h) and Willman~1 (14~h). For early
observations~\cite{Acciari2010, Aliu2015}, they used a simple event
counting analysis approach. More recently, they analyzed their full
datasets and combined them using advanced analysis
techniques~\cite{Archambault2017}.

\begin{figure}[t]
\centering
\includegraphics[width=1.0\textwidth]{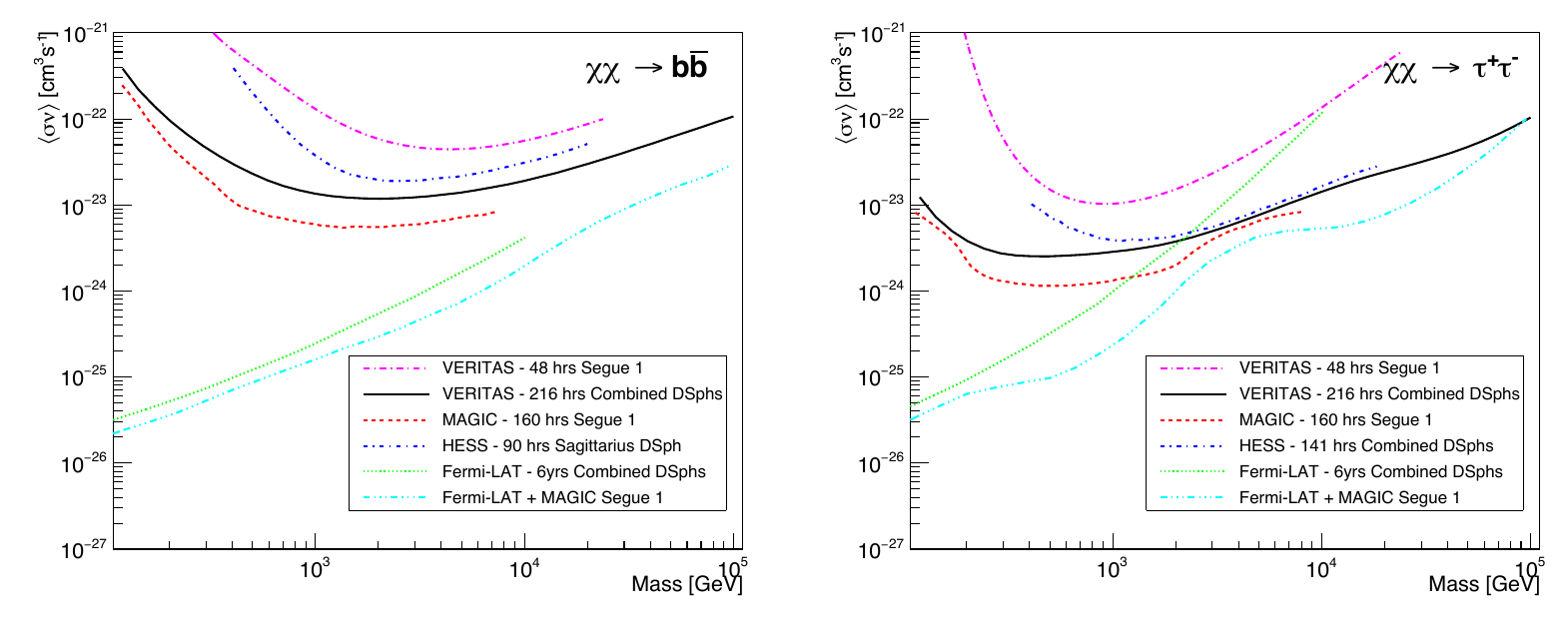}
\caption{The 95\% confidence-level upper limits to the dark matter
  annihilation cross-section into $\bb$ (\textbf{left}) and $\tautau$
  (\textbf{right}) pairs, obtained from dSph observations by VERITAS
  (black solid line), compared with results from other gamma-ray
  instruments (see legend for the details).  Reprinted figure with
  permission from reference \cite{Archambault2017}; copyright (2017)
  by the American Physical Society.}
\label{fig:VERITASlimits}
\end{figure}

In this latter work, the average gamma-ray spectra ($\dNdE$) for the
investigated dark matter annihilation channels were taken from the
PPPC 4 DM ID computation~\cite{ref:Cirelli2011}, and the differential
J-factors from Geringer-Sameth et al. (2014)~\cite{ref:Geringer2014}.
For the high-level, statistical data analysis, VERITAS used a test
statistic equivalent to the ratio of the following likelihood
function~\cite{ref:Geringer15}, namely:
\begin{equation} 
\lklgamma(\aaa \,|\,  \{E'_i,\theta'_i\}_{i=1,...,\Non}) = \prod_{i=1}^{\Non}
  f_{s+b}(E'_i,\theta'_i) \quad . \label{eq:veritaslkl} 
\end{equation}

This likelihood function is similar to the one used by MAGIC in the
Segue~1 analysis (Equation~(\ref{eq:magicunbinnedlkl})). They are both
unbinned simplified versions of the general likelihood function for
Cherenkov telescopes shown in Equation~(\ref{eq:onoffLkl}). With respect
to the MAGIC Segue~1 likelihood function, in
Equation~(\ref{eq:veritaslkl}) the external Poisson term for the total
number of observed events is omitted, and the event-wise term consists
in the evaluation of the {2-dimensional} PDF for the measured
energy $E'$ and the angular separation $\theta'$ between the measured
arrival direction and the dSph center. We remind the reader that
$f_{s+b} = \frac{1}{s+b}\left(s\, f_s+ b\,
  f_b\right)$. In the 2-dimensional case, assuming that the
convolution of the gamma-ray distribution with the IRF is radially
symmetric with respect to the center of the dSph (i.e.,  the dependence
on $\bp'$ reduces to a dependence on $\theta'$), then
$f_s(E',\theta')$ is given by:
\begin{equation}
  f_s(E',\theta') = \frac{2\pi \theta'  \Tobs}{s} 
  \int_0^\infty dE \int_{\domtot} d\Omega\,
 \frac{d^2\Phi}{dE\,d\Omega}\, \IRF(E',\bp'|E,\bp) \quad .
\label{eq:fsEtheta}
\end{equation}

Only events in an On region defined by a maximum distance of
$\theta'_\mathrm{cut}=0.17^\circ$ from the center of the dSphs are
considered, and the dependence of the effective area $\Aeff$ on the
arrival direction $\bp$ for events passing such cut is ignored. The
dependence of $f_b$ on $E'$ is modeled by smearing the distribution of
$E'$ measured for events of the background-control (Off) region,
whereas the spatial distribution is assumed to be uniform within the
On region. Both $b$ and $f_b$ are fixed during the likelihood
maximization, i.e.,  no statistical or systematic uncertainties in the
background estimation are considered. Moreover the J-factor uncertainty is
not included in the likelihood. Instead, the effect of the uncertainty
in $J$ is quantified by repeating the limit calculation over an
ensemble of dark matter halo realizations using, for each dSph, halo
parameter values randomly chosen from their inferred PDFs, and
reporting the 68\% confidence level containment quantiles of the
obtained distribution of results\footnote{\label{foot:xcdm} That is:
  limits, which are one-sided confidence intervals, are provided with
  error bars, which are two-sided confidence intervals. Some
  authors~\cite{ref:xkcd2110} have described graphically the
  potentially pernicious consequences of extending this
  practice.}. However, the main reported result in this case is still
the median of such distribution, which is only sensitive to the
central J-factor and not to its uncertainty, producing limits a factor
$\sim 2$ more constraining than if $\Jtot$ was considered a nuisance
parameter.

A possible advantage of the use of the likelihood function of
Equation~(\ref{eq:veritaslkl}) is that it allows a relatively simple
estimation of the PDF for the associated $-2\ln\lp$ test
statistic~\cite{ref:Geringer15} directly from the data and without
relying on the validity of the conditions of the Wilks' theorem. This
is so, because $-2\ln\lp$ can be expressed as the sum of two random
variables (those corresponding to the signal and background
contributions to $-2\ln\lp$, respectively), which, for the likelihood
function of Equation~(\ref{eq:veritaslkl}), are distributed according to
a {\em compound Poisson} distribution.
%Compound-Poisson-distributed
%variables are those that can be expressed as $\sum_{i=1}^N x_i$, with
%$N$ a Poisson-distributed random variable (with mean $\mu$) and $x_i$
%identically distributed independent random variables. The PDF of the
%sum of two random variables of known PDF is the convolution of the
%PDFs for each term, or, equivalently, the Fourier transform of the
%global PDF is the {product} of the Fourier transforms of the PDF
%for each term. The Fourier transform of the PDF for a compound Poisson
%distribution can be simply computed as a function of $\mu$ and the PDF
%for $x_i$. In our case, $N$ is the number of observed signal or
%background events, with $\mu = s$ or $b$, respectively, and the $x_i$
%are the single-event contributions to $-2\ln\lp$ from signal or
%background events, respectively. The distribution of single-event
%$-2\ln\lp$ values can be estimated by computing it in sufficiently
%fine bins of ($E'$,$\bp'$), and introducing the obtained values in a
%distribution weighted by their corresponding $s_{ij}$ value from
%Equation~(\ref{eq:sij}), for the case of signal, or the measured
%$b_{ij}$ from dedicated Off regions, for the case of background.
% Given that VERITAS results are obtained using the PDF for $-2\ln\lp$
%determined using this methodology (i.e.,  without relying on Wilks'
%theorem), 
VERITAS results are hence robust in the sense that have a well
determined confidence level under the assumption that the likelihood
function was correct.

VERITAS has not found evidence of dark matter signals from neither of
the four considered dSphs individually, or combined in a joint
analysis. The null-hypothesis significance is well within the
$\pm 2\sigma$ quantile, for all considered targets, annihilation
channels ($\uu$, $\dd$, $\ssbar$, $\bb$, $\ttbar$, $\ee $, $\mumu$,
$\tautau$, $\WW$, $ZZ$ and $hh$) and $\mdm$ values, except for
$\mdm \geq 5$~TeV dark matter particles annihilating into
$\gamma\gamma$ in Draco dSph. In this latter case, a {negative}
fluctuation slightly below $-2\sigma$ is observed, which is not
incompatible with purely statistical fluctuations, or could be
alternatively explained by unaccounted systematic uncertainties in the
background estimation. Figure~\ref{fig:VERITASlimits} shows VERITAS
limits to the annihilation cross-section into $\bb$ and $\tautau$
pairs, compared with other limits from dSph observations by other
gamma-ray instruments. The constraints reach
$\svnc \sim 10^{-23}$~\svunits\ at $\mdm \sim 1$~TeV for $\bb$, and
$\svnc \sim 3\times 10^{-24}$~\svunits\ at $\sim$300~GeV for $\tautau$
annihilation channels,~respectively.

\subsection {HAWC}

HAWC has searched for dark matter annihilation and decay signals in 15
dSphs observed during 507 days between November 2014 and June
2016~\cite{ref:Albert2018}. They computed the average gamma-ray
spectra per annihilation or decay event ($\dNdE$) using the PYTHIA
v8.2 simulation package~\cite{ref:pythia82}, and the J-factors using
the CLUMPY software package~\cite{ref:Clumpy}, assuming
NFW~\cite{ref:NFW1997} dark matter density profiles. The searches were
carried out using the binned likelihood function described in
Equation~(\ref{eq:binnedLkl}). Data were binned in
reconstructed energy $E'$ (referred to as $\fhit$ in HAWC
publications~\cite{ref:HAWCIRF}) covering the range between 500
GeV and 100 TeV, and in reconstructed
arrival direction $\bp'$, covering an area of $5^\circ$ radius around
each of the analyzed dSphs. The computation of
the signal events $s_{ij}$ in each bin was performed using Monte Carlo
simulations of the whole observations, assuming point-like sources and
a reference value of $\aaa$, and scaling the result for any other
needed value, which is equivalent to using Equation~(\ref{eq:sij}).

\begin{figure}[t]
\centering
\includegraphics[width=7.5 cm]{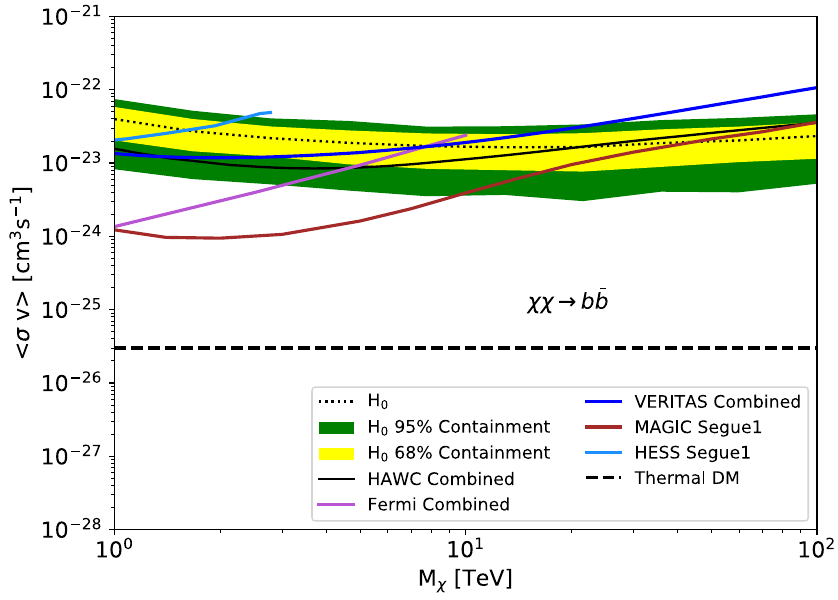}
\includegraphics[width=7.5 cm]{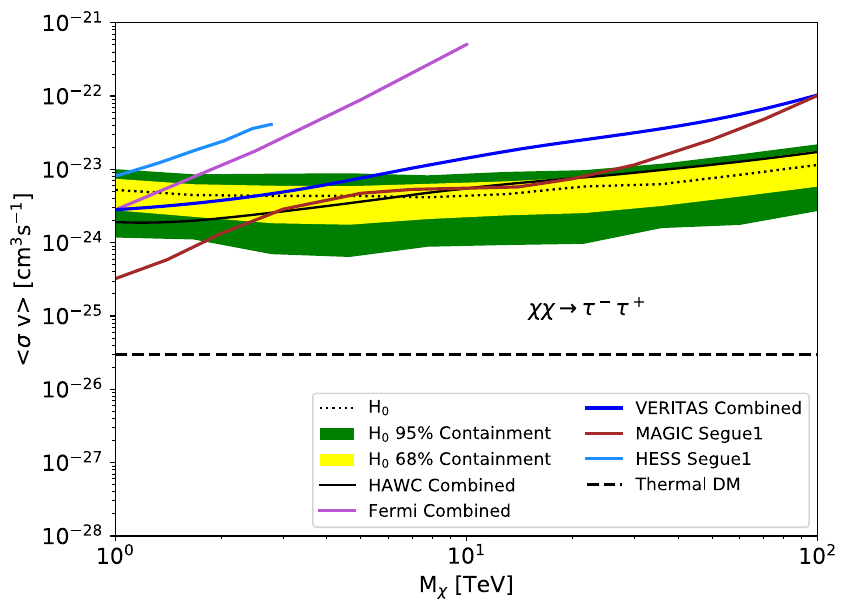}
\caption{The 95\% confidence level upper limits to the annihilation
  cross-section of dark matter particles annihilating into $\bb$
  (\textbf{left}) and $\tautau$ (\textbf{right}) pairs, from HAWC
  observations of dSphs (black solid line). Results from other
  gamma-ray instruments are also shown (see legend for details), as
  well as the median and 65\% and 95\% symmetric quantiles of the
  distribution of limits obtained under the null hypothesis. Figure
  reproduced with permission from reference~\cite{ref:Albert2018},
  \textcopyright AAS.}
\label{fig:HAWClimits}
\end{figure}

No nuisance parameters accounting for uncertainties in the background
estimation were considered, i.e.,  no $\lklmu$ term was included in the
Equation~(\ref{eq:binnedLkl}) likelihood function. The values $b_{ij}$
were estimated from the measured number of events in the same bin of
local (or detector) coordinates at times when such coordinates do not
correspond to any of the analyzed dSphs or any known HAWC
sources. Measured background rates at each local spatial bin were then
normalized using the all-sky event rate measured in 2-hour intervals.
Using this method, the statistics used for background estimation
correspond to an Off/On exposure ratio factor of
$\tau$ = 30--300~\cite{com:Harding2019}, and the related statistic
uncertainties (included in the case of Cherenkov telescopes by
the second Poisson term in Equation~(\ref{eq:onoffLkl})), can therefore
be safely neglected. However, the effect of the systematic uncertainty
associated to this method is not quantified or taken into account in
the analysis. In addition, similarly to the case of VERITAS, HAWC does
also not include in the maximum likelihood analysis the statistical
uncertainty in the J-factor, i.e.,  they ignore the $\lklJ$ term in
Equation~(\ref{eq:lklwithJ}). They do quantify the impact on the limits
caused by the consideration of the dSphs as point-like sources and by
several detector effects not perfectly under control in the Monte
Carlo simulations used for calibrating the detector.

HAWC has not found gamma-rays associated to dark matter annihilation
or decay from the examined dSphs, considered either individually or
collectively. The significance of rejection of the null hypothesis for
all considered targets, channels ($\bb$, $\ttbar$, $\tautau$, $\WW$
and $\mumu$), and $\mdm$ values (between 1 and 100 TeV) is within
$2\sigma$, except for few marginally larger negative fluctuations.
Figure~\ref{fig:HAWClimits} shows the limits to the annihilation cross
section obtained by HAWC for the $\bb$ and $\tautau$ annihilation
channels, compared to limits obtained by other gamma-ray instruments.
Limits reach $\svnc \sim 10^{-23}$~\svunits\ at $\mdm \sim 3$~TeV for
$\bb$, and $\svnc \sim 2\times 10^{-24}$~\svunits\ at $\sim$1~TeV for
$\tautau$ annihilation channels, respectively. For decay, lower limits
to the decay lifetime were set to $\taudmnc \sim 3\times 10^{26}$~s for
the 100 TeV mass dark matter particle decaying into $\bb$ pairs or
$\taudmnc \sim 10^{27}$~s for decaying into $\tautau$~pairs.

\subsection{Multi-Instrument Searches}

\begin{figure}[t]
\centering
\includegraphics[width=7.75 cm]{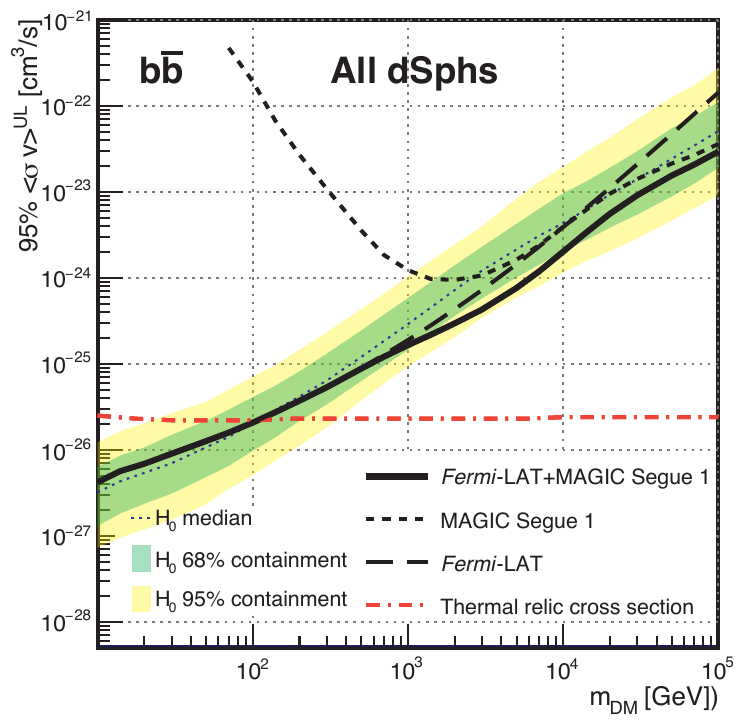}
\includegraphics[width=7.75 cm]{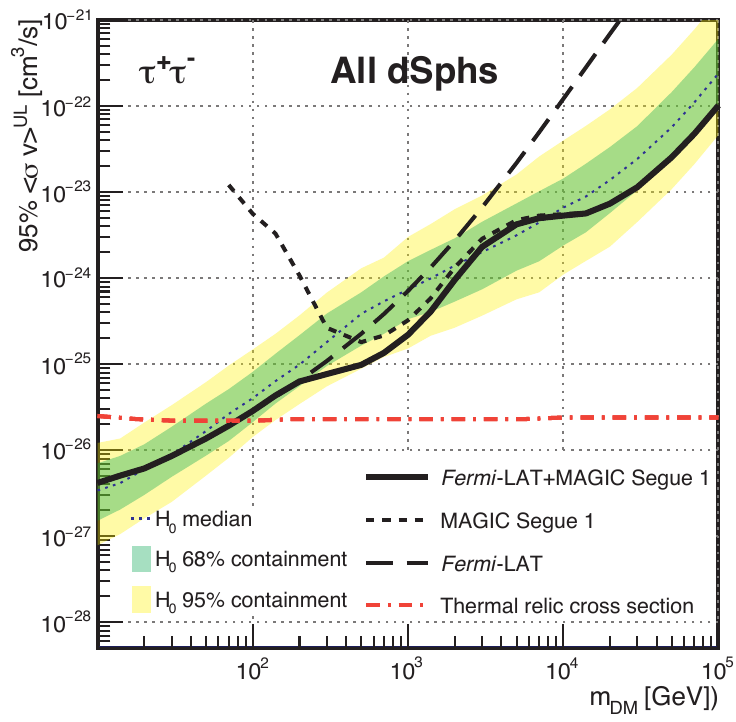}
\caption{The 95\% confidence level upper limits to the cross-section
  for dark matter particles annihilating into $\bb$ (\textbf{left})
  and $\tautau$ (\textbf{right}) pairs. Thick solid lines show the
  limits obtained by combining Fermi-LAT observations of 15 dSphs with
  MAGIC observations of Segue 1. Dashed lines show the limit obtained
  individually by MAGIC (short dashes) and Fermi-LAT (long dashes),
  respectively. The thin-dotted line, green and yellow bands show,
  respectively, the median and the two-sided 68\% and
  95\% symmetric quantiles for the distribution of limits under the
  null hypothesis. Reprinted figure with permission from
  reference~\cite{ref:MAGICLAT2016}, \textcopyright IOP Publishing Ltd
  and Sissa Medialab; reproduced by permission of IOP Publishing; all
  rights reserved.}
\label{fig:MAGIC-LAT}
\end{figure}

Following Equations~(\ref{eq:lklwithJ}) and (\ref{eq:multiinstrumentlkl}),
MAGIC and Fermi-LAT have computed a multi-target, multi-instrument,
joint likelihood, producing the first coherent joint search for
gamma-ray signals from annihilation of dark matter particles in the
mass range between 10~GeV and 100~TeV~\cite{ref:MAGICLAT2016}. The
data used in this work correspond to the Fermi-LAT
6-years~\cite{ref:Fermi2015} and the MAGIC
Segue~1~\cite{ref:MAGICSegue1} observations discussed earlier in
Sections~\ref{sec:LAT} and \ref{sec:MAGIC}, respectively. MAGIC
analysis was slightly adapted to match LAT conventions, in the
following aspects: (i) The determination of the J-factor; (ii) the
treatment of the statistical uncertainty of $\Jtot$ through the
$\lklJ$ term in Equation~(\ref{eq:lklwithJ}); and (iii) the treatment of
the cases in which the limits lie outside the physical ($\aaa\geq 0$)
region.

The MAGIC/Fermi-LAT combined search for dark matter did not produced a
positive signal, but it allowed setting global limits to the dark
matter annihilation cross section and, for the first time, a
meaningful comparison of the individual results obtained with the two
instruments. Figure~\ref{fig:MAGIC-LAT} shows the 95\% confidence
level limits to the cross-section of dark matter particles of mass in
the range between 10 GeV and 100 TeV annihilating into $\bb$ and
$\tautau$ pairs. The obtained limits are the currently most
constraining results from dSphs, and span the widest interval of
masses, covering the whole WIMP range. In the regions of mass where
Fermi-LAT and MAGIC achieve comparable sensitivities, the improvement
of the combined result with respect to those from individual
instruments reaches a factor $\sim 2$.

This approach is applicable to all the high-energy gamma-ray
instruments (and also to high energy neutrino telescopes, with slight
modifications in Equation~(\ref{eq:sij}) to account for the
oscillations). The so-called {Glory Duck} working group has
initiated an activity aimed at the combination of all dark matter
searches performed with Fermi-LAT, H.E.S.S, MAGIC, VERITAS
and HAWC using observations of dSphs~\cite{ref:GloryDuckICRC}. Each
collaboration will analyze their own datasets and will provide the
likelihood values as a function of the free parameter $\aaa$ (i.e.,\
the terms $\lklgammak$ in Equation~(\ref{eq:multiinstrumentlkl})) for
the different considered annihilation channels and $\mdm$ values, for
their combination and J-factor profiling through
Equation~(\ref{eq:lklwithJ}). Likelihood values from the different
instruments will be computed using the same conventions for the
computation of the gamma-ray spectra and the J-factors, as well as the
same statistical treatment of the data, most notably a common
consideration of all relevant uncertainties by the inclusion of the
corresponding nuisance parameters in the likelihood functions.
While in principle foreseen only for the combination of gamma-ray
data in the search of annihilation signals, this~work could pave the
path for other combined searches, such as searches for decay signals,
the inclusion of other kinds of targets or even extending the searches to
include also results from neutrino telescopes. This approach will
ensure that all the combined individual results will be directly
comparable among them, and will produce the legacy result of the dark
matter searches using the current generation of gamma-ray instruments.

\vspace{6pt}

%%%%%%%%%%%%%%%%%%%%%%%%%%%%%%%%%%%%%%%%%%
\funding{This work is partially funded by grant  FPA2017-87859-P from Ministerio de Econom\'{\i}a, Industria y Competitividad (Spain).}

%\acknowledgments{}

\conflictsofinterest{The author declares no conflict of interest.}

\reftitle{References}

% The following MDPI journals use author-date citation: Arts, Econometrics, Economies, Genealogy, Humanities, IJFS, JRFM, Laws, Religions, Risks, Social Sciences. For those journals, please follow the formatting guidelines on http://www.mdpi.com/authors/references
% To cite two works by the same author: \citeauthor{ref-journal-1a} (\citeyear{ref-journal-1a}, \citeyear{ref-journal-1b}). This produces: Whittaker (1967, 1975)
% To cite two works by the same author with specific pages: \citeauthor{ref-journal-3a} (\citeyear{ref-journal-3a}, p. 328; \citeyear{ref-journal-3b}, p.475). This produces: Wong (1999, p. 328; 2000, p. 475)

%=====================================
% References, variant B: external bibliography
%=====================================
%\externalbibliography{yes}
%\bibliography{your_external_BibTeX_file}

%%%%%%%%%%%%%%%%%%%%%%%%%%%%%%%%%%%%%%%%%%

%\appendixtitles{no} %Leave argument "no" if all appendix headings stay EMPTY (then no dot is printed after "Appendix A"). If the appendix sections contain a heading then change the argument to "yes".
%\appendixsections{multiple} %Leave argument "multiple" if there are multiple sections. Then a counter is printed ("Appendix A"). If there is only one appendix section then change the argument to "one" and no counter is printed ("Appendix").

%\appendix

\end{document}